\documentclass[twocolumn,trackchanges]{style/aastex631}

\usepackage{hyperref,enumerate}
\usepackage{xcolor}
\usepackage{rotating}

\hypersetup{
    unicode=false,                  
    pdftoolbar=true,                
    pdfmenubar=true,                
    pdffitwindow=true,              
    pdfstartview={FitH},            
    pdftitle={SPLUSJ1424-2542},     
    pdfauthor={vplacco},            
    pdfsubject={Astronomy},         
    pdfcreator={dvipdf},            
    pdfproducer={dvipdf},           
    pdfkeywords={metal-poor stars}, 
    pdfnewwindow=true,              
    colorlinks=true,                
    linkcolor=red,                  
    citecolor=blue,                 
    filecolor=magenta,              
    urlcolor=cyan,                  
    breaklinks=true,
    linktocpage
}

\newcommand{\vv}{{\tablenotemark{\footnotesize{a}}}}
\newcommand{\xx}{{\tablenotemark{\footnotesize{b}}}}

\accepted{to ApJ \today}
\begin{document}

\title{The \emph{R}-process Alliance: A Bright, Strongly \emph{R}-process-enhanced \\ Extremely Metal-poor Star Observed with GHOST} 
\shorttitle{G256353: A Bright $R$-process-enhanced Star Observed with Gemini-S/GHOST}

\shortauthors{Mardini et al.}
\correspondingauthor{Mohammad K.\ Mardini}
\email{mmardini@mit.edu}

\author[0000-0001-9178-3992]{Mohammad K.\ Mardini}
\affiliation{Department of Physics and Kavli Institute for Astrophysics and Space Research, Massachusetts Institute of Technology, Cambridge, MA 02139, USA}
\affiliation{Joint Institute for Nuclear Astrophysics, Center for the Evolution of the Elements (JINA-CEE), USA}

\author[0000-0002-2139-7145]{Anna Frebel}
\affiliation{Department of Physics and Kavli Institute for Astrophysics and Space Research, Massachusetts Institute of Technology, Cambridge, MA 02139, USA}
\affiliation{Joint Institute for Nuclear Astrophysics, Center for the Evolution of the Elements (JINA-CEE), USA}

\author[0000-0003-4479-1265]{Vinicius M.\ Placco}
\affiliation{NSF NOIRLab, Tucson, AZ 85719, USA}

\author[0000-0002-7155-679X]{Anirudh Chiti}
\affil{Department of Astronomy $\&$ Astrophysics, University of Chicago, 5640 S Ellis Avenue, Chicago, IL 60637, USA}
\affil{Kavli Institute for Cosmological Physics, University of Chicago, Chicago, IL 60637, USA}

\author[0000-0003-0864-9368]{Manolya Yatman}
\affiliation{Physics and Astronomy Department, Haverford College, PA, USA}

\author[0000-0003-4573-6233]{Timothy C.\ Beers}
\affiliation{Department of Physics and Astronomy, University of Notre Dame,
225 Nieuwland Science Hall, Notre Dame, IN 46556, USA}
\affiliation{Joint Institute for Nuclear Astrophysics, Center for the Evolution of the Elements (JINA-CEE), USA}

\author[0000-0002-8504-8470]{Rana Ezzeddine}
\affiliation{Department of Astronomy, University of Florida, Bryant Space Science Center, Gainesville, FL 32611, USA}
\affiliation{Joint Institute for Nuclear Astrophysics, Center for the Evolution of the Elements (JINA-CEE), USA}

\author[0000-0001-6154-8983]{Terese T.\ Hansen}
\affiliation{Department of Astronomy, Stockholm University, AlbaNova University Center, SE-106 91 Stockholm, Sweden}

\author[0000-0002-5463-6800]{Erika M.\ Holmbeck}
\affiliation{Lawrence Livermore National Laboratory, 7000 East Avenue, Livermore, CA 94550, USA}
\affiliation{Joint Institute for Nuclear Astrophysics, Center for the Evolution of the Elements (JINA-CEE), USA}

\author[0000-0001-5107-8930]{Ian U.\ Roederer}
\affiliation{Department of Physics and Astronomy, North Carolina State University,
2401 Stinson Dr, Box 8202, Raleigh, NC 27695, USA}
\affiliation{Joint Institute for Nuclear Astrophysics, Center for the Evolution of the Elements (JINA-CEE), USA}

\author[0000-0002-5095-4000]{Charli M.\ Sakari}
\affiliation{Department of Physics and Astronomy, San Francisco State University, San Francisco, CA 94132, USA}

\begin{abstract}
We present a detailed chemical-abundance and kinematic analysis of four extremely metal-poor (EMP; [Fe/H] $\leq -3.0$) stars identified from \textit{Gaia} BP/RP data in our ongoing search for the most primitive stars. This includes a primary target, \textit{Gaia}~DR3~2563539603865382656 (hereafter G256353), a strongly $r$-process-enhanced star with [Eu/Fe]~$= +1.20$ and [Ba/Eu]~$= -0.64$. Our results are based on high-resolution, high-signal-to-noise GHOST spectra from Gemini-South. For the full sample, we statistically match the light-element abundances with those predicted from Population\,III supernova models. The ``best-fit'' model suggests massive progenitors with stellar masses of M$_{\star}\sim$ 20-30\,M$_\odot$. In addition, we determine orbital histories for all of the stars. We find that Gaia~DR3~2887334237669844480 appears to be kinematically associated with Atari, an accreted structure in the Galactic disk. This star has low abundance ratios of strontium ([Sr/Fe] = $-$1.09) and barium ([Ba/Fe] = $-$0.37), which supports an accretion origin. For G256353, we determine chemical abundances for 15 neutron-capture elements. We compare the observed heavy-element pattern for G256353 with that of the Sun, HD~222925, and two neutron star merger models. The $r$-process elements in G256353 align reasonably well with HD~222925, the scaled-Solar pattern (except for the first peak), and a recent predicted pattern associated with neutron star mergers. This consistency reinforces the universality of the main $r$-process across diverse astrophysical environments.
\end{abstract}

\keywords{High resolution spectroscopy (2096), Chemical abundances (224), Nucleosynthesis (1131), $R$-process (1324), Milky Way dynamics (1051)}

\section{Introduction} \label{sec:intro}

The rapid neutron-capture (hereafter $r$-) process \citep{Burbidge1957,Cameron1957}, is a critical nucleosynthesis pathway that accounts for the production of approximately half of the heavy elements \mbox{(Z $\geq$ 30)}, including the actinides, in the Sun \citep{Beers2005,Frebel2018}. Identifying the astrophysical sites where the $r$-process occurs has been a significant challenge in astrophysics \citep[e.g.,][and references therein]{qian2004,Kasen2017,Cowan2021,Arcones2023,Bandyopadhyay2025}. Historically, core-collapse supernovae (CCSNe) were considered the primary candidates due to their extreme environments and high neutron flux \citep[e.g.,][]{,wanajo02,wanajo05,froehlich06,nomoto06,izutani09,farouqi10}. However, recent observations and simulations indicate that supernovae may not consistently achieve the neutron densities or eject sufficient neutron-rich material for robust $r$-process element formation, highlighting the need to consider alternative astrophysical sites \citep[e.g.,][]{Nishimura17,Fischer20,Ghosh22,Arcones2023}.

One of the most promising sites for $r$-process nucleosynthesis is neutron star (NS{\textendash}NS) mergers \citep[e.g.,][]{NSM_Lattimer,NSM_Eichler,Thielemann2017ARNPS}, which has been confirmed through gravitational wave observations, such as the remarkable detection of the event GW170817 \citep{Abbott_GW170817} and its subsequent kilonova AT2017gfo \citep{Abbott_AT2017gfo,Drout2017,Shappee2017}. These cataclysmic collisions between NS binaries form environments with extremely high neutron densities \mbox{($n_{n} > 10^{20}$\,cm$^{-3}$)}, ideal for the $r$-process. The ejected material from this merger has been observed to produce heavy elements \citep[][]{Watson_Sr_detection,Domoto_Sr_detection}. These observations support the idea that NS{\textendash}NS mergers play a significant role in cosmic $r$-process production. However, the relatively low frequency of NS{\textendash}NS collisions, combined with the expected long timescales required for NS binaries to merge \citep[see][and references therein]{Sana12}, raises questions about whether these events can fully account for all of the $r$-process elements observed in the Universe across different epochs \citep[see][and references therein]{cote19}.

In addition to NS{\textendash}NS mergers and CCSNe, other potential $r$-process sites include rare types of supernovae, such as magneto-rotational supernovae \citep[e.g.,][]{Symbalisty1985,Reichert2021}, where rapidly rotating, highly magnetized cores could provide the necessary conditions for $r$-process nucleosynthesis. Another potential environment for the $r$-process is in accretion disks associated with NS{\textendash}NS merger remnants or collapsars, where neutron-rich outflows may provide suitable conditions for heavy-element nucleosynthesis \citep[see][and references therein]{collapsars_1,collapsars_2}. It has also been suggested that, if the accretion disk around a neutron star enters the core of a massive evolved star, in the framework of a common-envelope jets supernova (CEJSN) $r$-process scenario, it could penetrate the crust of the neutron star, mix neutron-rich crust material into the disk, and enrich the jets that the disk launches with neutron-rich material (see \citealt{Soker2025}, and references therein).
 Despite these potential sites, each scenario faces limitations, such as uncertainties in the rate of occurrence, the amount of neutron-rich material ejected, and the exact conditions required for the $r$-process \citep[see][and references therein]{Cowan2021}. Thus, while significant progress has been made, further observational and theoretical work is necessary to fully understand the astrophysical origins of $r$-process elements.

In this context, the goal of the $R$-Process Alliance (RPA) has been to increase the number of known $r$-process-enhanced (RPE) stars in the Milky Way to better understand $r$-process nucleosynthesis and the associated astrophysical production site(s). Recent results from the RPA include detailed chemical abundances of both strongly enhanced $r$-II stars \mbox{([Eu/Fe$]> +0.7$)} and moderately enhanced \mbox{$r$-I} stars \mbox{($+0.3 <$ [Eu/Fe] $\leq +0.7$)} (\citealt{RPA_1st,RPA_2nd, RPA_3rd, RPA_4th,RPA_5th}). Building on these efforts, many individual RPE stars have been studied in detail to constrain $r$-process nucleosynthesis: J1521{\textendash}3538, an $r$-III star, with \mbox{[Eu/Fe] $> +2.0$}; \citealt{Cain20}), RAVE~J0949{\textendash}1617 (an $r+s$ star; \citealt{Maude18}), RAVE~J2038{\textendash}0023 (U detection; \citealt{UII_Placco}), S-PLUS~J1424{\textendash}2542;  an (actinide-boost star; \citealt{Placco2023}), J0954$+$5246; (U detection; \citealt{UII_Holmbeck}), J2213{\textendash}5137 (an $r$-III star with a U detection; \citealt{Roederer24high_eu}), and a sample of stars that exhibit potential abundance signatures of transuranic fission fragments \citep{roederer23_fission}.

Here we present an additional effort to identify metal-poor stars with $r$-process enhancements selected from the Gaia BP/RP (also known as XP) survey \citep{Gaia_DR3}, and observed with the GHOST spectrograph on the Gemini-South telescope. As part of this first sample of high-resolution spectra, we include observations of the bright EMP $r$-II star Gaia DR3 2563539603865382656 (hereafter G256353), originally identified by \citet{Li_2022}, with \mbox{$T_{\rm{eff}} = 4784$\,K,  [Fe/H] $= -3.22$} and [Eu/Fe] = +1.35. In this paper, we determine abundances for 12 additional elements, including eight neutron-capture elements, and present the full chemo{\textendash}dynamical analysis of this star, as well as that of three newly discovered halo EMP stars in our sample. 

The remainder of this paper is organized as follows.  Section~\ref{sec:obs} describes the target selection and observations using the GHOST spectrograph. The stellar parameters and chemical abundances are derived in Section~\ref{sec:params_abund}. Section~\ref{sec:kinem} details the dynamical histories of our sample using a time-dependent Galactic potential. We discuss our chemodynamic results in Section~\ref{sec:diss}, and summarize our findings in Section~\ref{sec:conc}. 

\section{Target selection and Observations}\label{sec:obs}

The Gaia Data Processing and Analysis Consortium provided the {\texttt{GaiaXPy }}\footnote{We used GaiaXPy 1.1.3: Available at \url{https://gaia-dpci.github.io/GaiaXPy-website}} package, which allows users to estimate the expected flux of objects with BP/RP spectra using pre-loaded transmission curves. 
Currently, the package includes medium- and broad-band (e.g., $u$, $i$) and narrow-band (e.g., SkyMapper \ion{Ca}{2}~K) photometric filters. In particular, the SkyMapper $v$ filter, covering the 3600{\textendash}4100\,\AA\, range, and the Pristine filter, with an even narrower 3900{\textendash}4000\,\AA\, range, both encompass the metallicity sensitive \ion{Ca}{2}~K line at $\sim$ 3933\,\AA, which is a key feature for determining [Fe/H]. The combination of estimating fluxes from the broad- and narrow-band filters for Gaia XP observations can then be efficiently used to robustly identify metal-poor candidates \citep[see][]{Pristine_gaiaxp,Mardini24b,Chiti24_nat,Ou25}.

We used the filters mentioned above to derive photometric [Fe/H] estimates for a sample of $\sim$ seven million Gaia XP sources. These estimates were derived by matching precomputed synthetic photometry grids with the corresponding observed values, following the methodology described in \citet{Chiti_SMSS_cat2021}. This procedure was further refined to incorporate the Pristine \ion{Ca}{2}~K filter and a machine-learning gradient-boosted tree model. This combined approach yielded three independent [Fe/H] estimates for each candidate. In general, our Gaia XP [Fe/H] estimates exhibit good agreement, compared to high-resolution spectroscopic [Fe/H] values, with differences within 0.30\,dex, and extending down to [Fe/H] $\sim-4.0$. This accuracy is comparable to the typical uncertainties of medium{\textendash}resolution spectroscopic results, which generally have uncertainties of $\sim 0.20$--0.25\,dex. A more detailed explanation of this methodology, along with a catalog of metallicity estimates, will be provided in a forthcoming paper (M. Mardini et al. 2026, in preparation).

We selected four candidates with [Fe/H]$_{SMSS} \leq -3.0$, [Fe/H]$_{Pris} \leq -3.0$, and [Fe/H]$_{ML} \leq -3.0$ from our catalog as part of our effort to identify the most metal-poor stars in the Milky Way. Given that these filter sets produce generally higher estimated [Fe/H] estimates when high carbon enhancement is present (as described by \citealt{pristine_EMP}), we do not expect our EMP candidates to exhibit strong carbon enhancement. Table~\ref{tab:obs} lists the Gaia {\tt source\_id} of these candidates. 

\begin{deluxetable*}{lrrrrl}
\tablecaption{Observational Properties of Our Sample Stars\label{tab:obs}}
\tabletypesize{\tiny}
\tablewidth{0pc}
\tablehead{
\colhead{} &
\colhead{G256353} &
\colhead{G288733} &
\colhead{G297027} &
\colhead{G470812} }
\startdata
Gaia DR3 Source\_ID & 2563539603865382656 & 2887334237669844480  & 2970270979572570240  &   4708124561256755328              \\ 
Other names & 2MASS~J01240263+0458154 &2MASS~J05471221$-$3609160  & 2MASS~J05400394$-$1837487  &   SMSS~J003608.15$-$650438.9      \\ 
Right ascension [HMS]&                    ~~01:24:02.63& ~~05:47:12.22&  ~~05:40:03.95& ~~00:36:08.16                        \\ 
Declination [DMS]&                       $+$04:58:15.24&$-$36:09:15.99& $-$18:37:48.91& $-$65:04:38.97                       \\ 
Date of observation&         2023 Dec 12          &2023 Dec 12    &2023 Dec 11     &2023 Dec 12                              \\ 
Exp. time [s] &                   2160               &840           &900            &1200                                    \\ 
Galactic longitude [deg] &                  222.492     &   241.542    &   222.493     &   305.548                           \\ 
Galactic latitude [deg]&                 $-$23.834     &$-$27.891     & $-$23.834     & $-$51.973                            \\ 
Parallax [mas]&                       ~ ~ ~0.1853      &~ ~ ~0.3119   &~ ~ ~ ~0.6391  &~ ~ ~0.0983                           \\ 
Parallax\_error [mas]&                   ~ ~ ~0.0162   &~ ~ ~0.0109   &~ ~ ~ ~0.0168  &~ ~ ~0.0106                           \\ 
Distance [kpc] &                          5.44          &  3.22        & 1.57          &  9.90                               \\ 
Proper motion in RA [mas~y$^{-1}$]&  ~~$-$1.7955   &~ ~ ~ ~2.67623&~ ~ ~ ~8.1189  &  ~~$-$1.4579                             \\ 
Proper motion in DEC [mas~y$^{-1}$]&      ~~$-$6.3438   &~ ~ ~ ~2.73729&~~~~~$-$9.3723 &  ~~$-$4.7970                        \\ 
$G$ magnitude [mag]&                12.23              &~10.40        &~~~11.05       &  ~~12.27                             \\ 
$BP$ magnitude [mag]&                12.71             &~10.99        &~~~11.49       &  ~~12.91                             \\ 
$RP$ magnitude [mag]&                11.58             &~ 9.67        &~~~10.43       &  ~~11.51                             \\ 
Reddening E(B $-$ V) [mag]&                  ~ ~ 0.0233        &~ ~ ~0.0451   &~ ~ ~~~0.0527  &~ ~ ~~0.0146                  \\ 
Bolometric correction [mag]&~~$-$0.38                &~ ~$-$0.48  &~ ~ ~$-$0.32 &~~~~$-$0.59                                 \\ 
Signal-to-noise ratio at $\sim$4000\AA\ per pixel &  50              &45            &       33      &       38                                        \\ 
Effective Temperature [K]&           4810            &4510          &5050           &       4376                             \\ 
Log of surface gravity [log(cm~s$^{-2}$)] &          1.49            &1.00          &2.16           &       0.76             \\ 
Microturbulent velocity [km~s$^{-1}$]&         2.03            &2.23          &1.66           &       2.55                   \\ 
Metallicity &                  $-$3.22            &$-3.01$       &$-3.08$        &       $-3.14$                             \\ 
Radial velocity (RV) [km~s$^{-1}$] &             $-174.0$           &167.9        &320.6         &       69.6                 \\ 
RV from \citet{Gaia_DR3}                &             $-170.3$           &169.2        &322.1         &       69.8               \\ 
RV from \citet{Aoki22}                &             $-173.7$           &\nodata        &\nodata        &       \nodata                              \\
\enddata
\end{deluxetable*}

We then obtained high{\textendash}resolution spectroscopic data for these candidates. The spectra were obtained with the newly commissioned Gemini High-resolution Optical SpecTrograph (GHOST; \citealt{ireland2014,mcconnachie2022,hayes2023,kalari2024}) at the Gemini-South telescope. The observations were conducted as part of the 2023B semester GHOST shared-risk call for proposals (Program ID: GS-2023B-FT-301\footnote{\href{https://archive.gemini.edu/searchform/GS-2023B-FT-301}{https://archive.gemini.edu/searchform/GS-2023B-FT-301}}). The chosen instrument setup was the standard resolution of R~$\sim47,000$ and target mode \texttt{IFU1:Target$|$IFU2:Sky}, which provides improved sky subtraction. All spectra were taken with a 2 $\times$ 4 binning (spectral $\times$ spatial). The wavelength coverage is 3600 to 5400\,{\AA} for the blue camera and 5250 to 10600\,{\AA} for the red camera.

The reduced spectra were provided by the US National Gemini Office \citep[US NGO;][]{placco2024}. Data reduction was performed using v3.1 of the {\texttt{DRAGONS}}\footnote{\href{https://dragons.readthedocs.io/en/stable/}{https://dragons.readthedocs.io/en/stable/}.} software package \citep{labrie2023}. The reduction steps included bias and flat-field corrections, wavelength calibration, barycentric correction, sky subtraction, extraction of individual orders, and variance-weighted stitching of the spectral orders. The final 1D spectra created by {\texttt{DRAGONS}} were written in text format using standard \texttt{NOIRLab IRAF}\footnote{NOIRLab IRAF is distributed by the Community Science and Data Center at NSF NOIRLab, which is managed by the Association of Universities for Research in Astronomy (AURA) under a cooperative agreement with the U.S. National Science Foundation.} routines. Finally, we combined the two back-to-back individual exposures for each program star using a simple average without applying any rejection criteria.

We calculated the heliocentric radial velocity (RV) by cross-correlating our GHOST final combined spectra with a rest-frame Magellan/MIKE spectrum of HD~122563, a red giant star with [Fe/H] $\sim -2.8$ \citep{Aoki2025} as no RV standard star was observed as part of our program. The cross-correlation was performed around the Ca\,triplet region, which is a strong feature commonly used in radial velocity measurements for metal-poor stars. Table~\ref{tab:obs} lists the observing details for our four targets. Astrometric information, magnitudes, and parallaxes are adopted from \citet{Gaia_DR3}. Reddening and bolometric corrections are taken from \citet{Casagrande2018BC1} and \citet{schlafly2011}, respectively. The distances are calculated as in \citet{Mardini2022}. We also included RV values reported in Gaia DR3 \citep{Gaia_DR3} and \citet{Aoki22}. None of the stars exhibit significant RV variations, suggesting that our target stars are not likely members of binary systems.

\section{Equivalent widths, Stellar Parameters, and Chemical Abundances}\label{sec:params_abund}

\subsection{Equivalent Widths}\label{sec:ews}

We used the linelist from \citet{Roederer18_lines} along with atomic, hyperfine-structure, isotopic splitting \citep[see][]{sneden_araa}, and molecular line data from the \texttt{linemake} code\footnote{Available at \url{https://github.com/vmplacco/linemake}} \citep{Placco2021_linemake}. The list features isolated lines, free from blends or other contamination, especially those caused by molecular CH band features.

\begin{deluxetable*}{lrrrrcrcrcrcrr}[!htbp]
\tabletypesize{\tiny}
\tabletypesize{\footnotesize}
\tablewidth{0pc}
\tablecaption{\label{tab:eqw} Atomic Data, Equivalent Widths, and Derived Abundances}
\tablehead{
\colhead{species}&
\colhead{$\lambda$}&
\colhead{$\chi$} &
\colhead{$\log\,gf$}&
\multicolumn{2}{c}{G256353} &
\multicolumn{2}{c}{G288733} &
\multicolumn{2}{c}{G297027} &
\multicolumn{2}{c}{G470812} &\\
\colhead{}&
\colhead{{(\AA})}&
\colhead{(eV)}&
\colhead{}&
\colhead{$EW$\,(m{\AA})}&
\colhead{$\log\epsilon$\,(X)}&
\colhead{$EW$\,(m{\AA})}&
\colhead{$\log\epsilon$\,(X)}&
\colhead{$EW$\,(m{\AA})}&
\colhead{$\log\epsilon$\,(X)}&
\colhead{$EW$\,(m{\AA})}&
\colhead{$\log\epsilon$\,(X)}&
\colhead{Ref}&
}
\startdata
CH             & 4313.00 & \nodata & \nodata & \nodata     &    4.65 & \nodata     &    4.22 & \nodata     &    5.62 & \nodata     &  4.54    & (1)\\
\ion{Na}{1}    & 5889.95 & 0.00    &    0.11 & 152.42  &    3.56 & 197.38  &    3.73 & 126.07  &    3.49 & 195.09  &  3.40                    & (5)\\
\ion{Na}{1}    & 5895.92 & 0.00    & $-$0.19 & 125.85  &    3.39 & 172.61  &    3.69 & 101.60  &    3.31 & 166.49  &  3.32                    & (5)\\
\ion{Mg}{1}    & 3829.35 & 2.71    & $-$0.23 & 149.63  &    4.83 & \nodata & \nodata & \nodata & \nodata & 189.19  &  4.86                    & (5)\\
\ion{Mg}{1}    & 4057.51 & 4.35    & $-$0.90 &  22.60  &    4.93 & 36.61   &    5.04 & \nodata & \nodata &  27.67  &  4.80                    & (5)\\
\ion{Mg}{1}    & 4167.27 & 4.35    & $-$0.74 &  28.92  &    4.90 & 42.87   &    4.97 & 25.80   &    4.96 &  45.35  &  4.95                    & (5)\\
\ion{Mg}{1}    & 4702.99 & 4.33    & $-$0.44 &  47.11  &    4.87 & 65.23   &    4.96 & 37.04   &    4.83 &  71.29  &  4.97                    & (5)\\
\ion{Mg}{1}    & 5172.68 & 2.71    & $-$0.36 & 167.96  &    4.86 & 212.15  &    4.99 & 149.17  &    4.84 & 225.93  &  4.90                    & (6)\\
\ion{Mg}{1}    & 5183.60 & 2.72    & $-$0.17 & 188.43  &    4.92 & 237.80  &    5.02 & 164.36  &    4.84 & 248.77  &  4.90                    & (6)\\
\enddata
\tablecomments{This table is available in its entirety in machine-readable form. Relevant references for atomic data are presented in Appendix A.}
\end{deluxetable*}

For the equivalent width (EW) analysis, we employed the Spectroscopy Made Hard ({\texttt{SMHr}}) software\footnote{Available at \url{https://github.com/andycasey/smhr}} \citep{casey14}, which is specifically designed to facilitate the measurement of absorption lines in stellar spectra. We fitted the absorption features in our high-resolution spectrum with Gaussian profiles to obtain the EW measurements of the detectable lines listed in Table~\ref{tab:eqw}. Finally, each line was visually examined to ensure good continuum placement and fitting of the line. 

\begin{deluxetable*}{lrrrrr r rrrrr r rrrrr r rrrrr}[h!]
\tabletypesize{\tiny}
\tablewidth{0pc}
\tablecaption{Chemical Abundances for our Target Stars Determined from the Gemini-S/GHOST Spectra \label{tab:abund}}
\tablehead{
\colhead{Species} &
\multicolumn{5}{c}{G256353} & \colhead{} & \multicolumn{5}{c}{G288733} & \colhead{} & \multicolumn{5}{c}{G297027} & \colhead{} & \multicolumn{5}{c}{G470812} \\
\cline{2-6} \cline{8-12} \cline{14-18} \cline{20-24}
\colhead{} &
\colhead{N} & \colhead{$\log{\epsilon}$} & \colhead{$\sigma$\vv} & \colhead{[X/H]} & \colhead{[X/Fe]} & \colhead{} &
\colhead{N} & \colhead{$\log{\epsilon}$} & \colhead{$\sigma$} & \colhead{[X/H]} & \colhead{[X/Fe]} & \colhead{} &
\colhead{N} & \colhead{$\log{\epsilon}$} & \colhead{$\sigma$} & \colhead{[X/H]} & \colhead{[X/Fe]} & \colhead{} &
\colhead{N} & \colhead{$\log{\epsilon}$} & \colhead{$\sigma$} & \colhead{[X/H]} & \colhead{[X/Fe]} 
} 
\startdata
C        &  2 &  4.65 & 0.05 & $-$3.78 & $-$0.58 & &  2 &  4.22 & 0.05 & $-$4.21 & $-$1.23 & &  2 &  5.62 & 0.06 & $-$2.80 & +0.26 & &  2 &  4.54 & 0.05 & $-$3.88 & $-$0.75 \\
C\xx     & \nodata & \nodata & \nodata & $-$3.37 & $-$0.17 & & \nodata & \nodata & \nodata & $-$3.45 & $-$0.47 & & \nodata & \nodata & \nodata & $-$2.61 & +0.45 & & \nodata & \nodata & \nodata & $-$3.13 & 0.00 \\
Na \,I   &  2 &  3.47 & 0.06 & $-$2.77 & +0.43 & &  3 &  3.71 & 0.05 & $-$2.53 & +0.45 & &  2 &  3.40 & 0.07 & $-$2.84 & +0.22 & &  2 &  3.36 & 0.05 & $-$2.88 & +0.25 \\
Mg \,I   &  8 &  4.92 & 0.05 & $-$2.68 & +0.52 & &  7 &  5.05 & 0.05 & $-$2.55 & +0.43 & &  6 &  4.90 & 0.05 & $-$2.70 & +0.36 & &  7 &  4.92 & 0.05 & $-$2.68 & +0.45 \\
Al \,I   &  2 &  2.80 & 0.10 & $-$3.65 & $-$0.45 & &  2 &  2.81 & 0.05 & $-$3.63 & $-$0.65 & &  2 &  2.95 & 0.05 & $-$3.50 & $-$0.44 & &  2 &  2.97 & 0.05 & $-$3.48 & $-$0.35 \\
Si \,I   &  2 &  4.99 & 0.05 & $-$2.52 & +0.68 & &  2 &  4.83 & 0.05 & $-$2.68 & +0.30 & &  2 &  5.23 & 0.05 & $-$2.28 & +0.78 & &  2 &  4.95 & 0.06 & $-$2.56 & +0.57 \\
Ca \,I   & 13 &  3.58 & 0.05 & $-$2.76 & +0.44 & & 18 &  3.71 & 0.05 & $-$2.63 & +0.35 & & 19 &  3.63 & 0.05 & $-$2.71 & +0.35 & & 19 &  3.52 & 0.05 & $-$2.82 & +0.31 \\
Sc \,II  & 10 &  0.14 & 0.05 & $-$3.01 & +0.19 & & 10 &  0.06 & 0.05 & $-$3.08 & $-$0.10 & &  9 &  0.22 & 0.05 & $-$2.92 & +0.14 & &  8 &  0.10 & 0.05 & $-$3.04 & +0.09 \\
Ti \,I   & 13 &  2.11 & 0.05 & $-$2.83 & +0.37 & & 18 &  2.25 & 0.05 & $-$2.69 & +0.29 & & 13 &  2.16 & 0.05 & $-$2.78 & +0.28 & & 14 &  1.86 & 0.05 & $-$3.08 & +0.05 \\
Ti \,II  & 26 &  2.22 & 0.05 & $-$2.73 & +0.47 & & 21 &  2.55 & 0.05 & $-$2.40 & +0.58 & & 25 &  2.25 & 0.05 & $-$2.70 & +0.36 & & 24 &  2.13 & 0.05 & $-$2.82 & +0.31 \\
V \,II   &  2 &  0.84 & 0.05 & $-$3.08 & +0.12 & &  2 &  0.92 & 0.09 & $-$3.01 & $-$0.03 & &  1 &  1.14 & 0.10 & $-$2.79 & +0.27 & &  2 &  0.79 & 0.07 & $-$3.13 & 0.00 \\
Cr \,I   &  7 &  2.26 & 0.05 & $-$3.38 & $-$0.18 & &  8 &  2.43 & 0.05 & $-$3.21 & $-$0.23 & &  6 &  2.27 & 0.05 & $-$3.36 & $-$0.30 & &  7 &  2.31 & 0.05 & $-$3.32 & $-$0.19 \\
Mn \,I   &  4 &  1.57 & 0.05 & $-$3.86 & $-$0.66 & &  4 &  1.83 & 0.06 & $-$3.59 & $-$0.61 & &  5 &  1.91 & 0.05 & $-$3.51 & $-$0.45 & &  6 &  1.79 & 0.05 & $-$3.63 & $-$0.50 \\
Fe \,I   &122 &  4.30 & 0.05 & $-$3.20 &  0.00 & &136 &  4.52 & 0.05 & $-$2.98 &  0.00 & &136 &  4.44 & 0.05 & $-$3.06 &  0.00 & &125 &  4.37 & 0.05 & $-$3.13 &  0.00 \\
Fe \,II  & 13 &  4.37 & 0.05 & $-$3.13 & +0.07 & & 11 &  4.60 & 0.05 & $-$2.90 & +0.08 & & 11 &  4.50 & 0.05 & $-$3.00 & +0.06 & & 11 &  4.40 & 0.05 & $-$3.10 & +0.03 \\
Co \,I   &  4 &  1.86 & 0.05 & $-$3.13 & +0.07 & &  5 &  2.09 & 0.05 & $-$2.89 & +0.09 & &  5 &  2.16 & 0.05 & $-$2.83 & +0.23 & &  5 &  1.84 & 0.05 & $-$3.15 & $-$0.02 \\
Ni \,I   &  7 &  3.07 & 0.05 & $-$3.15 & +0.05 & & 10 &  3.32 & 0.05 & $-$2.90 & +0.08 & &  7 &  3.23 & 0.05 & $-$2.99 & +0.07 & & 11 &  3.15 & 0.05 & $-$3.07 & +0.06 \\
Zn \,I   &  3 &  1.79 & 0.05 & $-$2.77 & +0.43 & &  3 &  1.89 & 0.06 & $-$2.66 & +0.32 & &  3 &  2.21 & 0.09 & $-$2.35 & +0.71 & &  3 &  1.61 & 0.06 & $-$2.95 & +0.18 \\
Sr \,II  &  2 &  0.17 & 0.05 & $-$2.70 & +0.50 & &  2 & $-$1.21 & 0.05 & $-$4.07 & $-$1.09 & &  2 & $-$0.11 & 0.05 & $-$2.97 & +0.09 & &  2 & $-$0.31 & 0.05 & $-$3.18 & $-$0.05 \\
Y \,II   &  6 & $-$0.66 & 0.07 & $-$2.87 & +0.33 & & \nodata & \nodata & \nodata & \nodata & \nodata & & \nodata & \nodata & \nodata & \nodata & \nodata & & \nodata & \nodata & \nodata & \nodata & \nodata \\
Zr \,II  &  6 &  0.03 & 0.05 & $-$2.54 & +0.66 & & \nodata & \nodata & \nodata & \nodata & \nodata & & \nodata & \nodata & \nodata & \nodata & \nodata & & \nodata & \nodata & \nodata & \nodata & \nodata \\
Ba \,II  &  3 & $-$0.46 & 0.05 & $-$2.64 & +0.56 & &  3 & $-$1.24 & 0.05 & $-$3.42 & $-$0.44 & &  3 & $-$1.20 & 0.05 & $-$3.37 & $-$0.31 & &  3 & $-$1.55 & 0.05 & $-$3.73 & $-$0.60 \\
La \,II  &  2 & $-$1.09 & 0.11 & $-$2.19 & +1.01 & & \nodata & \nodata & \nodata & \nodata & \nodata & & \nodata & \nodata & \nodata & \nodata & \nodata & & \nodata & \nodata & \nodata & \nodata & \nodata \\
Ce \,II  &  4 & $-$0.96 & 0.05 & $-$2.34 & +0.66 & & \nodata & \nodata & \nodata & \nodata & \nodata & & \nodata & \nodata & \nodata & \nodata & \nodata & & \nodata & \nodata & \nodata & \nodata & \nodata \\
Pr \,II  &  2 & $-$1.44 & 0.05 & $-$2.16 & +1.04 & & \nodata & \nodata & \nodata & \nodata & \nodata & & \nodata & \nodata & \nodata & \nodata & \nodata & & \nodata & \nodata & \nodata & \nodata & \nodata \\
Nd \,II  &  5 & $-$0.85 & 0.09 & $-$2.27 & +0.93 & & \nodata & \nodata & \nodata & \nodata & \nodata & & \nodata & \nodata & \nodata & \nodata & \nodata & & \nodata & \nodata & \nodata & \nodata & \nodata \\
Sm \,II  &  8 & $-$1.01 & 0.06 & $-$1.96 & +1.24 & & \nodata & \nodata & \nodata & \nodata & \nodata & & \nodata & \nodata & \nodata & \nodata & \nodata & & \nodata & \nodata & \nodata & \nodata & \nodata \\
Eu \,II  &  3 & $-$1.49 & 0.05 & $-$2.00 & +1.20 & & \nodata & \nodata & \nodata & \nodata & \nodata & & \nodata & \nodata & \nodata & \nodata & \nodata & & \nodata & \nodata & \nodata & \nodata & \nodata \\
Gd \,II  &  3 & $-$0.75 & 0.17 & $-$1.82 & +1.38 & & \nodata & \nodata & \nodata & \nodata & \nodata & & \nodata & \nodata & \nodata & \nodata & \nodata & & \nodata & \nodata & \nodata & \nodata & \nodata \\
Tb \,II  &  1 & $<$-$1.16$ & \nodata & $<$-$2.26$ & $<$+0.94 & & \nodata & \nodata & \nodata & \nodata & \nodata & & \nodata & \nodata & \nodata & \nodata & \nodata & & \nodata & \nodata & \nodata & \nodata & \nodata \\
Dy \,II  &  3 & $-$0.87 & 0.05 & $-$1.97 & +1.23 & & \nodata & \nodata & \nodata & \nodata & \nodata & & \nodata & \nodata & \nodata & \nodata & \nodata & & \nodata & \nodata & \nodata & \nodata & \nodata \\
Ho \,II  &  3 & $-$1.61 & 0.06 & $-$2.09 & +1.11 & & \nodata & \nodata & \nodata & \nodata & \nodata & & \nodata & \nodata & \nodata & \nodata & \nodata & & \nodata & \nodata & \nodata & \nodata & \nodata \\
Er \,II  &  2 & $-$1.11 & 0.05 & $-$2.03 & +1.17 & & \nodata & \nodata & \nodata & \nodata & \nodata & & \nodata & \nodata & \nodata & \nodata & \nodata & & \nodata & \nodata & \nodata & \nodata & \nodata \\
\enddata
\tablenotetext{a}{The uncertainties reported here are statistical only, while an example of the corresponding systematic uncertainties are provided in Table~\ref{tab:sys}.}
\tablenotetext{b}{Calculated based on procedures presented in \citet{placco14_carbon}.}
\end{deluxetable*}

\subsection{Atmospheric Parameters}\label{sec:params}

To determine the atmospheric stellar parameters for our target stars, we utilized the RPA methodology, which combines multiple observational sources, including magnitudes, distance, and iron-line abundances.

Photometric effective temperature ($T_{\rm{eff}}$) estimates are based on the precise $G$, $BP$, and $RP$ magnitudes reported in the Gaia Data Release 3 (DR3) catalog \citep{Gaia_DR3}. We applied the bolometric correction methods described by \citet{Casagrande2018BC1} and \citet{Casagrande2018BC2} to obtain accurate intrinsic magnitudes. This step is essential for ensuring that our temperature estimates reflect the true stellar luminosity. We then used the best polynomial fit of the color–$T_{\rm{eff}}$ relation from \citet{Mucciarelli2021}. By leveraging the empirical data and theoretical models of this polynomial fit, we translate each star's observed color indices into an effective temperature.

surface gravity ($\log g$) estimates for our target stars used parallax measurements from Gaia DR3 \citep{Gaia_DR3}. To account for systematic offsets in the Gaia data, parallax corrections were applied following the routines described in \citet{Lindegren_Parallax_2021}. We computed the heliocentric distances as presented in \citet{Mardini2022}. We then used fundamental relations that correlate luminosity, mass, and radius to derive surface gravity \citep[see equations 2 and 3 in][]{Mardini_2019a}.

Finally, we used the latest version of the MOOG spectral synthesis code \citep{moog}\footnote{Available at \url{https://github.com/alexji/moog17scat} and wrapped within {\texttt{SMHr}}}. We used the one-dimensional plane-parallel model atmospheres with $\alpha$-enhancement from \citet{Castelli2004}. We then used our measured EWs to obtain the abundances of the \ion{Fe}{1} and \ion{Fe}{2} lines for our final abundance [Fe/H], which we obtained from the \ion{Fe}{1} lines. We also required no trend between the individual measurements of these lines to iteratively constrain the microturbulence (v$_{\xi}$). Our adopted stellar parameters are listed in Table~\ref{tab:obs}. We adopted 2$\sigma$ uncertainties for the effective temperature derived from pure Gaia colors, corresponding to $\Delta T_{\rm eff}=100$\,K \citep[see][and references therein]{Mucciarelli2021}. In addition, we assumed typical uncertainties of $\sigma \log(g)=0.3$\,dex and $\sigma v_{\xi}=0.3$\,km,s$^{-1}$.

\begin{figure*}
\begin{center}
\includegraphics[clip=true,width=\textwidth]{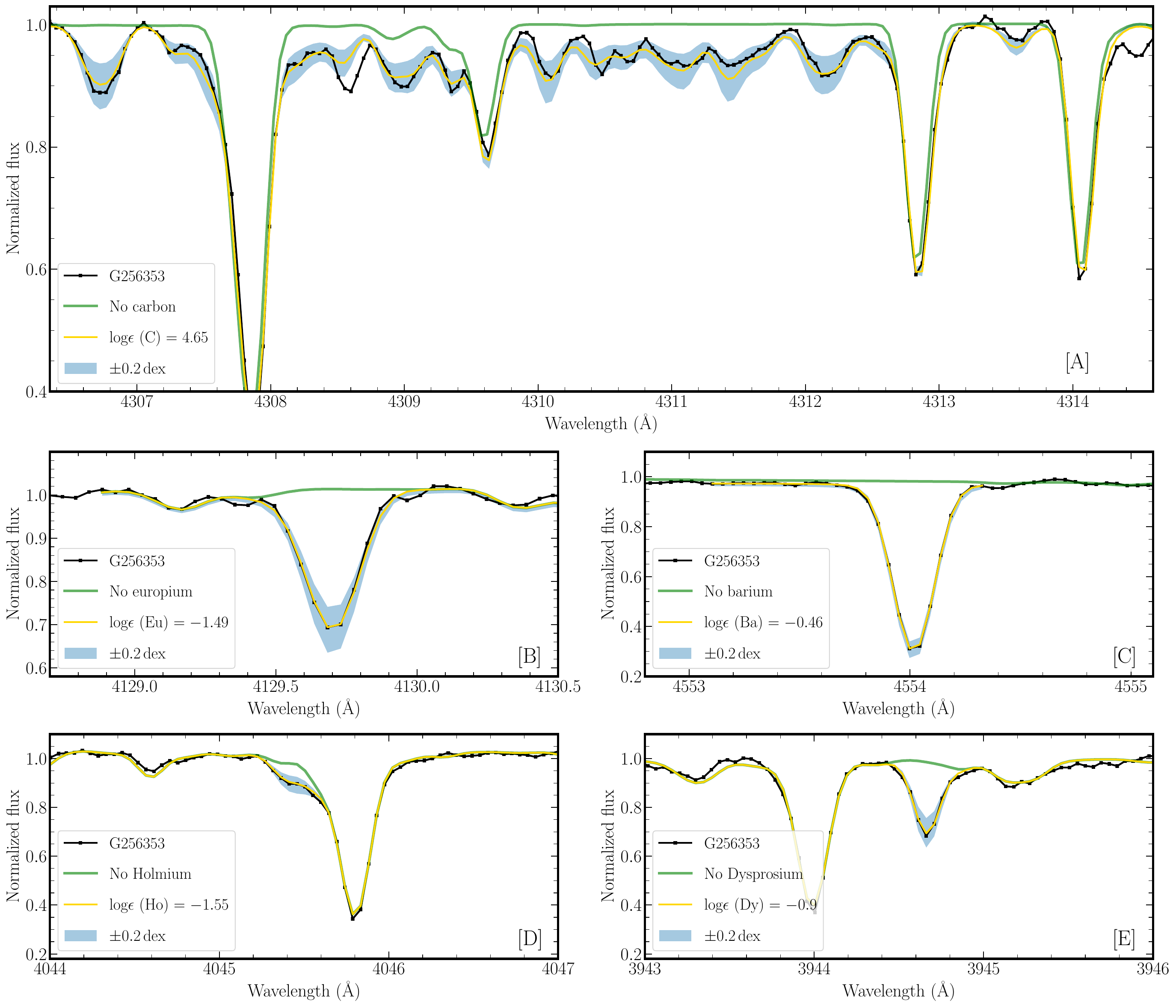} 
\caption{Portions of the spectrum of G256353 used to determine the chemical abundances of key elements. Filled squares denote the spectrum, while gold lines represent the best-fit spectrum syntheses. Green lines indicate syntheses without contributions from the relevant element. The shaded regions reflect $\pm 0.2$\,dex uncertainties. Each legend shows the derived abundances for the corresponding lines. \label{fig:synth}}
\end{center}
\end{figure*}

\begin{figure*}[!htbp]
\begin{center}
\includegraphics[clip=true,width=\textwidth]{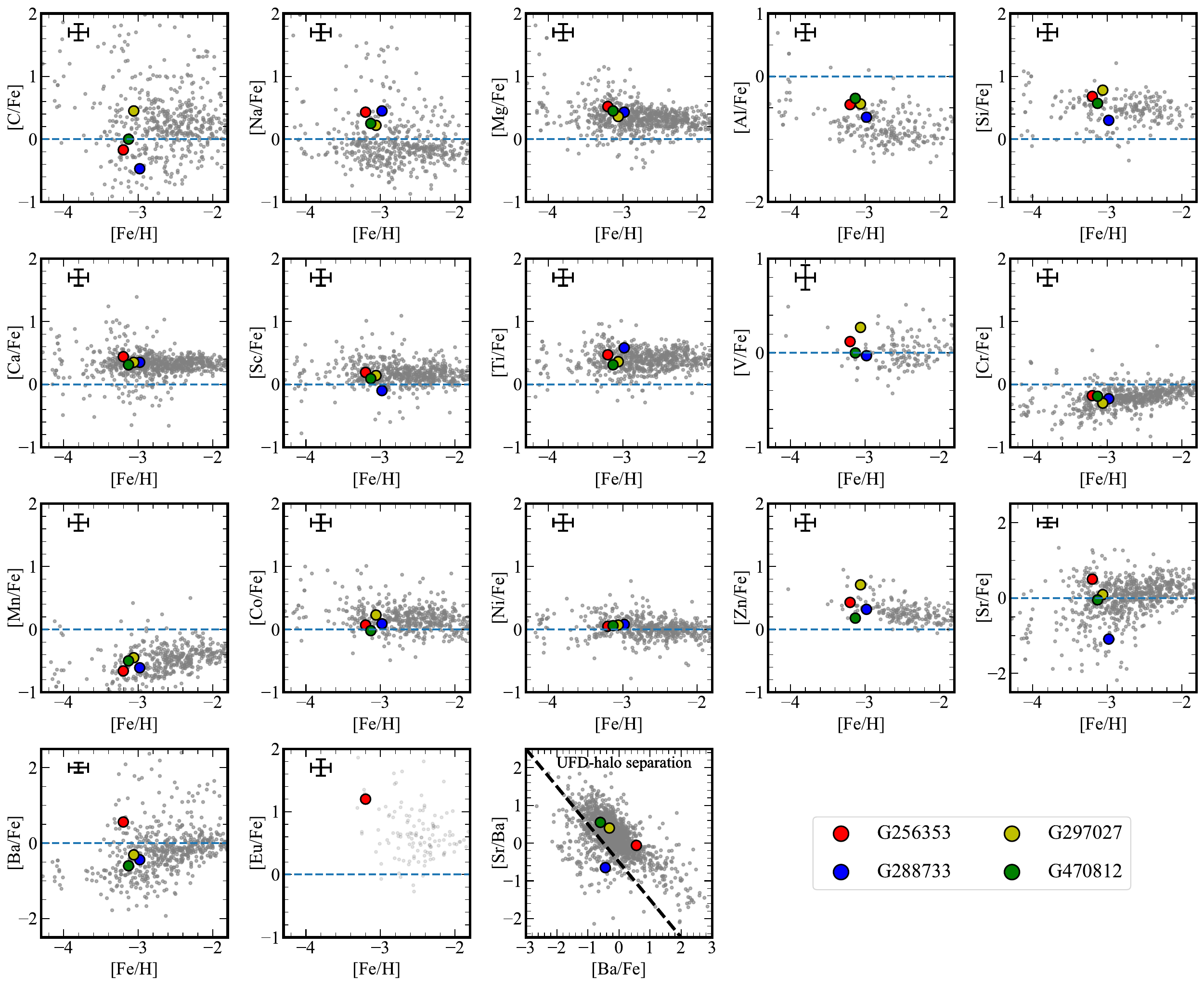} 
 \caption{Observed chemical abundances of our target stars (represented by filled circles) are compared with those of other metal-poor halo stars from the JINAbase (shown as gray points, relevant references are listed in Appendix B.). Overall, the abundances of our stars align well with those of the other metal-poor stars. In the ``UFD–halo separation'' panel, the dashed line indicates the relation [Sr/Ba] = $-$[Ba/Fe] $-$ 0.50, which delineates the region occupied by UFD stars from that of other Galactic metal-poor stars \citep[e.g.,][]{Mardini24a}. G288733 appears to fall in the UFD region. The legend displays the Gaia DR source ID for our target stars. Error bars illustrate the representative uncertainties ($\pm 0.15$\,dex) for each element in both our sample and the JINAbase data set. 
 \label{fig:abund}}
\end{center}
\end{figure*}

\subsection{Chemical Abundances and Associated Uncertainties}\label{sec:abund}

Owing to the fairly good signal-to-noise (S/N) ratio of the GHOST spectra, we derived chemical abundances for 17 elements in all target stars. In G256353, which exhibits $r$-process enhancement, we additionally measured 13 neutron-capture elements. To determine these abundances, we employed both EW and spectral synthesis analysis. We adopted the Solar photospheric abundances from \citet{asplund09} as a reference. The final abundance ratios ([X/Fe]), the number of absorption lines used, and the standard errors are listed in Table~\ref{tab:abund}.

\subsubsection{Light Elements}

To determine accurate carbon abundances, we utilized the molecular CH $G${\textendash}band head located at 4313\,{\AA}, as illustrated in the upper panel of Figure~\ref{fig:synth}. We also used the CH features around 4217\,{\AA} to estimate the $^{12}$C/$^{13}$C ratios of the target stars, as described in \citet{Mardini_2019b}. We then corrected the derived carbon abundances based on the procedures described in \citet{placco14_carbon} to account for the evolutionary status of each star. We find very low [C/Fe] ratios for three of our stars, except for \mbox{G297027}, which exhibits a moderate carbon enhancement of \mbox{[C/Fe] = +0.45}. 

In addition to carbon, we derived the chemical abundances of sodium, magnesium, aluminum, silicon, calcium, scandium, titanium, vanadium, chromium, manganese, cobalt, nickel, and zinc. These elements provide insights into the nucleosynthetic processes that occurred prior to the formation of our target stars. See \citet{Arcones2023} for more context on the nucleosynthesis processes in supernova explosions that are responsible for the production of these light elements in the early universe.

Figure~\ref{fig:abund} shows the derived abundance ([X/Fe]) ratios as a function of [Fe/H]. Gray points are literature values collected by JINAbase\footnote{Compilation available at~\url{https://jinabase.pythonanywhere.com}} \citep{jinabase}. The abundances observed in our target stars reinforce the idea that EMP stars formed from a medium pre-enriched in light elements. This consistency suggests that the enrichment of the interstellar medium occurred through similar nucleosynthetic processes, producing consistent quantities and distributions of light elements across these stars. This homogeneity points to a shared origin or set of processes, such as early supernova events or other enrichment mechanisms, which played a key role in shaping the chemical composition of the earliest stars in our Galaxy.

\subsubsection{Neutron-capture Elements}

Very few neutron-capture elements are detected in the spectra for three of our target stars. This is primarily because the spectral lines of these elements are generally very weak, making them challenging to detect and analyze. We consistently detected two Sr II lines at wavelengths 4077\,{\AA} and 4215\,{\AA}, and three Ba II lines at 5853\,{\AA}, 6141\,{\AA}, and 6497\,{\AA}. These lines were found across the spectra of our target stars. The neutron-capture elements strontium and barium, together with europium, serve as tracers of the contributions to the $s$- and $r$-process in stars, respectively. 

One exception in our sample is G256353, which exhibits a strong enhancement in $r$-process elements. Consequently, we measured an additional 13 neutron-capture elements in its spectrum. These elements are products of the $r$-process, indicating that this star likely experienced a distinct or more pronounced $r$-process enrichment compared to the other stars in our sample. The lower panels of Figure~\ref{fig:synth} present the spectral synthesis used to determine the barium and europium abundances for G256353. We incorporated hyperfine-structure effects from {\texttt{linemake}}\,\citep{Placco2021_linemake,Placco2021ascl}, isotopic shifts, and HFS broadening, where applicable, using $r$-process isotopic ratios from \citet{sneden_araa}.

The overall uniformity of abundances observed among the light elements of Galactic halo stars does not extend to the heavy elements, which exhibit significant star-to-star scatter, especially at the lowest metallicities \citep[e.g.,][]{Frebel2018}. The significant element-to-element scatter among heavy neutron-capture elements thus suggests that the processes responsible for these elements may have occurred in a more stochastic fashion, which would leave the gas inhomogeneously mixed. Alternatively, individual nucleosynthesis events may have contributed disproportionately to the overall distribution of these elements within a given gas cloud. This contrasts with the more uniform enrichment mechanism(s) of lighter elements.

\begin{deluxetable}{lccccc}
\tabletypesize{\small}
\tabletypesize{\footnotesize}
\tablewidth{0pc}
\tablecaption{Example Systematic Abundance Uncertainties for \mbox{G256353} 
\label{tab:sys}}
\tablehead{
\colhead{Elem}&
\colhead{$\Delta$\ensuremath{T_\mathrm{eff}}}&
\colhead{$\Delta$\ensuremath{\log\,g}}&
\colhead{$\Delta\xi$}&
\colhead{$\sigma$}&
\colhead{$\sigma_{\rm tot}$}\\
\colhead{}&
\colhead{$+$100\,K}&
\colhead{$+$0.3 dex}&
\colhead{$+$0.3 km~s$^{-1}$}&
\colhead{(dex)}&
\colhead{(dex)}}
\startdata
\ion{Na}{1} & $+$0.12 & $-$0.03 & $-$0.10 & 0.06 & 0.17 \\
\ion{Mg}{1} & $+$0.14 & $-$0.05 & $-$0.09 & 0.09 & 0.20 \\
\ion{Ca}{1} & $+$0.09 & $-$0.03 & $-$0.04 & 0.08 & 0.13 \\
\ion{Sc}{2} & $+$0.07 & $+$0.03 & $-$0.06 & 0.04 & 0.10 \\
\ion{Ti}{2} & $+$0.10 & $+$0.05 & $-$0.05 & 0.08 & 0.15 \\
\ion{V}{2}  & $+$0.08 & $+$0.09 & $-$0.03 & 0.01 & 0.12 \\
\ion{Cr}{1} & $+$0.16 & $+$0.03 & $-$0.06 & 0.09 & 0.20 \\
\ion{Mn}{1} & $+$0.11 & $-$0.03 & $-$0.09 & 0.08 & 0.17 \\
\ion{Co}{1} & $+$0.14 & $-$0.05 & $-$0.11 & 0.05 & 0.19 \\
\ion{Ni}{1} & $+$0.16 & $-$0.04 & $-$0.06 & 0.14 & 0.22 \\
\ion{Zn}{1} & $+$0.05 & $+$0.01 & $-$0.02 & 0.05 & 0.07 \\
\hline
\hline
\ion{Sr}{2} & $+$0.07 & $-$0.05 & $-$0.04 & 0.05 & 0.11 \\
\ion{Y}{2}  & $+$0.04 & $-$0.06 & $-$0.05 & 0.02 & 0.09 \\
\ion{Zr}{2} & $+$0.05 & $+$0.06 & $-$0.08 & 0.02 & 0.11 \\
\ion{Ba}{2} & $+$0.10 & $-$0.08 & $-$0.09 & 0.02 & 0.16 \\
\ion{La}{2} & $+$0.09 & $-$0.06 & $-$0.07 & 0.02 & 0.13 \\
\ion{Ce}{2} & $+$0.06 & $+$0.07 & $-$0.06 & 0.02 & 0.11 \\
\ion{Pr}{2} & $+$0.07 & $-$0.05 & $-$0.05 & 0.02 & 0.10 \\
\ion{Nd}{2} & $+$0.07 & $+$0.09 & $-$0.08 & 0.01 & 0.14 \\
\ion{Sm}{2} & $+$0.06 & $+$0.06 & $-$0.07 & 0.01 & 0.11 \\
\ion{Eu}{2} & $+$0.09 & $-$0.06 & $-$0.05 & 0.02 & 0.12 \\
\ion{Gd}{2} & $+$0.08 & $-$0.09 & $-$0.08 & 0.04 & 0.15 \\
\ion{Dy}{2} & $+$0.06 & $+$0.09 & $-$0.07 & 0.02 & 0.13 \\
\ion{Ho}{2} & $+$0.09 & $-$0.08 & $-$0.07 & 0.04 & 0.14 \\
\ion{Er}{2} & $+$0.10 & $-$0.09 & $-$0.06 & 0.04 & 0.15 \\
\enddata
\end{deluxetable}

\subsubsection{Associated Uncertainties}

To ensure the precision and reliability of our results, we undertook a detailed quantification of the systematic uncertainties that arise from variations in the atmospheric parameters, particularly for the light elements. We systematically changed each atmospheric parameter with its adopted uncertainty (see Section~\ref{sec:params}), and recalculated the abundances for light elements. This allowed us to identify how sensitive our derived abundances are to variations in each parameter. Table~\ref{tab:sys} lists the abundance variations when changing the atmospheric parameters. The $\sigma$ values represent the standard error of the mean value. The $\sigma_{\rm tot}$ values represent the quadratic sum of the individual error estimates. 

The data presented in Table~\ref{tab:sys} strongly support the precision of our abundance measurements. The error estimates for our abundance values are relatively small, indicating that these uncertainties do not significantly affect the overall results. The minimal variation in the error estimates further reinforces the precision of our analysis and the quality of the collected spectra.

We also note that, given the metal-poor nature of our targets ([Fe/H] $\leq -3$), departures from local thermodynamic equilibrium (LTE) are expected to affect the formation of several spectral lines, including those of iron. In particular, Fe\,I lines are known to be susceptible to NLTE over-ionization in metal-poor atmospheres. For stars with [Fe/H] $\sim-3$, \citet{ezzeddine17} calculated $<0.25$\,dex corrections. Similar NLTE effects have been found by \citet{Mashonkina11}, \citet{bergemann12}, and \citet{lind12}. While our analysis is performed under the LTE assumption, these effects may introduce non-negligible uncertainties. A detailed NLTE treatment is beyond the scope of this work and might be explored in future studies.

\begin{deluxetable*}{lcccrccrrrr}
\tabletypesize{\tiny}
\tabletypesize{\footnotesize}
\tablewidth{0pc}
\tablecaption{Dynamical Parameters for our Target Stars\label{tab:kine}}
\tablehead{
\colhead{Star}&
\colhead{r$_\ensuremath{\mathrm{apo}}$}&
\colhead{r$_\ensuremath{\mathrm{peri}}$}&
\colhead{Z$_{max}$}&
\colhead{V$_\phi$}&
\colhead{ecc}&
\colhead{E}&
\colhead{J$_\phi$}&
\colhead{J$_r$}&
\colhead{J$_z$}&
\colhead{Membership}\\
\colhead{}&
\colhead{(kpc)}&
\colhead{(kpc)}&
\colhead{(kpc)}&
\colhead{(km s$^{-1}$)}&
\colhead{}&
\colhead{(10$^{5}$\,km$^{2}$ s$^{-1}$)}&
\colhead{(kpc km\,s$^{-1}$)}&
\colhead{(kpc km\,s$^{-1}$)}&
\colhead{(kpc km\,s$^{-1}$)}&
\colhead{}\\
}
\startdata
G256353  &15.51$_{0.98}^{0.95}$&2.85$_{0.04}^{0.09}$& 6.23$_{0.28}^{1.01}$&60.52$_{0.98}^{0.95}$&0.73$_{0.02}^{0.01}$&$-$1.41$_{0.02}^{0.02}$&$-$1152.84$_{20.29}^{14.06}$  &824.49$_{28.59}^{26.2}$ & 164.67$_{9.89}^{6.43}$   &  Halo       \\
G288733  &11.74$_{0.09}^{0.09}$&3.83$_{0.04}^{0.05}$& 1.97$_{0.02}^{0.03}$ &$-27.15$$_{0.58}^{0.65}$&0.22$_{0.01}^{0.01}$&$-$1.63$_{0.01}^{0.01}$&$-$1209.49$_{16.33}^{14.81}$  &372.69$_{1.12}^{1.53}$ & 36.24$_{0.43}^{0.51}$   &  Atari       \\
G297027  &12.51$_{0.02}^{0.03}$&0.84$_{0.05}^{0.05}$& 9.61$_{0.04}^{0.10}$ &$-95.1$$_{0.35}^{0.30}$&0.87$_{0.01}^{0.01}$&$-$1.58$_{0.01}^{0.01}$&$+$182.06$_{15.49}^{14.05}$&1319.59$_{7.99}^{8.1}$& 164.78$_{1.59}^{2.41}$   &  Halo       \\
G470812  &14.50$_{1.62}^{1.38}$&1.76$_{1.09}^{0.78}$&13.25$_{0.92}^{0.98}$ &94.29$_{0.40}^{0.45}$&0.71$_{0.04}^{0.08}$&$-$1.44$_{0.06}^{0.05}$&$+$421.94$_{70.51}^{68.36}$&669.82$_{29.81}^{23.39}$ & 642.04$_{30.93}^{30.74}$   &  Halo       \\
\enddata
\tablecomments{We define J$_\phi$ as $-$L$_z$}
\end{deluxetable*}

\section{Orbital histories of target stars}\label{sec:kinem}

\begin{figure*}[!htbp]                   
\begin{center}
\includegraphics[clip=true,width=\textwidth]{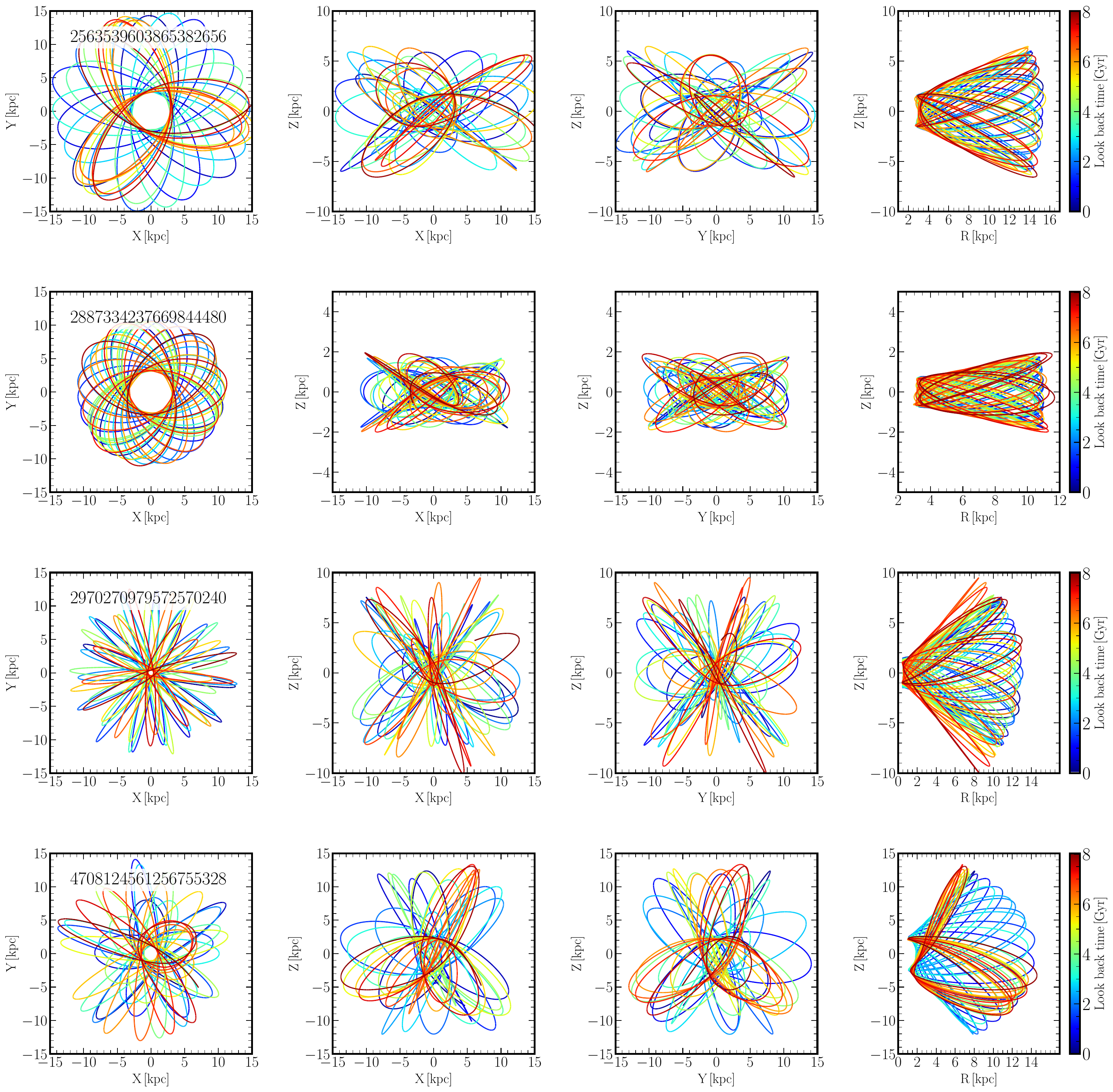}
\caption{Orbital histories for our target stars: G256353 (first row), G288733 (second row), \mbox{G297027} (third row), and \mbox{G470812} (fourth row). The halo-like kinematics are clearly seen except for G288733 (second row), which exhibits an Atari-like orbital history that keeps the star confined to the disk system \citep{Mardini2022}. }
\label{fig:orbits}
\end{center}
\end{figure*}

Combining orbital information (e.g., velocity, eccentricity, and inclination) with the results from our chemical-abundance analysis helps to discern the cosmic origins of our stars. A full kinematic analysis thus allows associating a given star with a distinct stellar population \citep[e.g.,][]{GSE_Helmi18,GSE_Belokurov18,Zepeda23,Mardini2022,SASS_stars,Hong24,Peter2024} to learn if it was formed in-situ within the Milky Way or it was accreted from a satellite galaxy \citep[see][and references therein]{Helmi_review20}.

We collected astrometric data for our target stars from Gaia DR3 \citep{Gaia_DR3}, which includes proper motions in both directions ($\mu_{\alpha}\cos\delta$, $\mu_{\delta}$) and parallaxes ($\varpi$). We adopted our own line-of-sight velocities (see Table~\ref{tab:obs}). We corrected $\varpi$ as suggested by \citet{Lindegren_Parallax_2021}, ensuring the accuracy of our distance measurements (see Table~\ref{tab:obs}). Subsequently, we derived the heliocentric distances of our target stars by applying an exponentially decreasing space density prior, as outlined in \citet{Mardini2022}. Combining these parameters with positions (RA and Dec) allows us to perform a comprehensive three-dimensional orbital history of our stars.

We used \texttt{The-ORIENT}\footnote{\url{https://github.com/Mohammad-Mardini/The-ORIENT}} \citep{Mardini_2020,Mardini2022b} to calculate the apocentric (r$_\ensuremath{\mathrm{apo}}$) and pericentric radii (r$_\ensuremath{\mathrm{peri}}$) for the stars in our sample, and determined the maximum offsets from the Galactic midplane (Z$_\ensuremath{\mathrm{max}}$). Additionally, we calculated the orbital eccentricity, defined as e = (r$_\ensuremath{\mathrm{apo}}$ $-$ r$_\ensuremath{\mathrm{peri}}$) / (r$_\ensuremath{\mathrm{apo}}$ + r$_\ensuremath{\mathrm{peri}}$). To estimate the uncertainties in our calculations, we performed a Monte Carlo simulation by generating 10,000 realizations for each input parameter of the astrometric data, drawing from Gaussian distributions. From these 10,000 realizations, we computed the mean value as our final result. To reflect the natural dispersion of our calculations, the uncertainties were quantified using the 16th and 84th percentiles, which serve as the lower and upper bounds, respectively. Table~\ref{tab:kine} lists these parameters and uncertainties.

Using our action-velocity separation tool \citep[detailed in][]{Mardini2022}, we calculated the relative likelihood that our stars belong to a particular population. The detailed mathematical framework underlying this is described further in \citet{Mardini2022}. The resulting memberships of our target stars are listed in Table~\ref{tab:kine}. Figure~\ref{fig:orbits} shows the orbital trajectories for our stars over the past 8\,Gyr and supports our membership assignments. Our sample stars, with the exception of G288733, belong to the Galactic halo. This origin is further supported by their large excursions from the Galactic plane, which suggest that they were formed in a more chaotic environment and possibly influenced by past mergers and accretion events.

\section{Discussion}\label{sec:diss}

In the following, we discuss the origin of the light elements \mbox{(Z $\leq 30$)} observed in our target stars. Additionally, the distinct chemical abundances and orbital parameters of G256353 and G288733 separate them from the rest of our sample, suggesting that they experienced different formation environments and/or evolutionary processes. We explore possible hypotheses to explain the origin of these stars, including scenarios such as accretion from satellite galaxies and potential enrichment processes.

\begin{figure*}
\begin{center}
\includegraphics[clip=true,width=\textwidth]{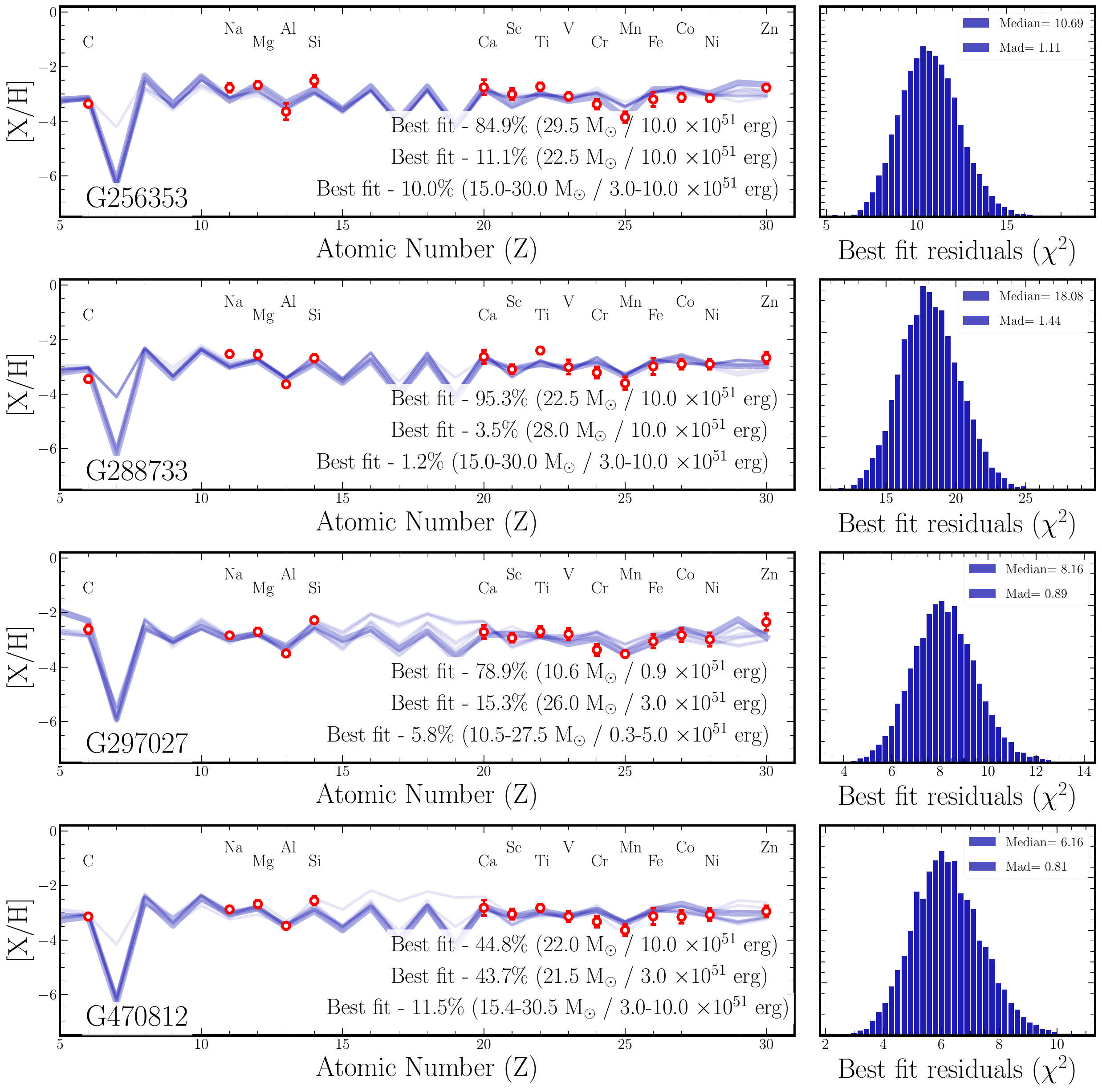} 
\caption{The left panels show a direct comparison between the observed light-element abundances and the theoretical predictions from \citet{heger_woosley10}. The right panels display the median and mean of the residuals, which illustrate the spread of our ``best-fit'' models. The majority of these models converged around a mass of  $\sim$20\,M$_{\odot}$. \label{fig:starfit}}
\end{center}
\end{figure*}

\subsection{Origin of the Light Elements}

We attempted to determine the primary characteristics of the progenitor population responsible for enriching the gas clouds that formed our target stars with light elements. To do this, we compare the observed abundances with theoretical nucleosynthesis yields from Population\,III supernova models \citep{heger_woosley10}. We employed fallback supernova models with stellar masses ranging from 10 to 100\,M$_\odot$ and explosion energies ranging from 0.3 $\times$ 10$^{51}$ to 10.0 $\times$ 10$^{51}$\,erg. To find the best-fit model that minimizes the differences between our observed abundances and those of the theoretical yields, we utilized the $\chi^{2}$ matching algorithm known as {\texttt{StarFit}}\footnote{\url{http://starfit.org}}.

We statistically compared each theoretical model with 10,000 realizations of the observed abundance pattern for each star in our sample. These realizations were generated using Gaussian distributions based on the observed abundances and their corresponding uncertainties. This approach enabled us to identify the progenitor model that best reproduces the light-element abundances observed in our target stars. The left panels of Figure~\ref{fig:starfit} show the predicted yields of the ``best-fit'' models compared to the observed abundances. The legends show the key parameters of each ``best-fit'' model. 

The ``best-fit'' model for G256353, which matched 84.9\% of the generated patterns, has a stellar mass of 29.5\,M$_\odot$ and explosion energy of 10.0 $\times$ 10$^{51}$\,erg. Another possible model with slightly lower stellar mass \mbox{(M = 22.5\,M$_\odot$)} and similar explosion energy matched 11.1\% of the generated patterns. Furthermore, we identified a set of 59 models, which span a broad parameter space, as potential fits for the remaining generated patterns.

The algorithm {\texttt{StarFit}} suggested a ``best-fit'' model for G288733, which successfully matched 95.3\% of the abundance patterns generated. This model has stellar mass of 22.5\,M$_\odot$ and an explosion energy of \mbox{10.0 $\times$ 10$^{51}$\,erg}. In addition to this dominant model, we found another possible progenitor with a slightly higher stellar mass of 28.0\,M$_\odot$ but similar explosion energy, which matched 3.5\% of the generated patterns. Beyond these two models, we identified eight additional models that could potentially fit the remaining generated patterns. These models have various stellar masses \mbox{(15-30\,M$_\odot$)} and explosion energies (3.0-10.0 $\times$ 10$^{51}$\,erg).

We found that 78.9\% of the abundance patterns generated for G297027 closely matched the predicted yields of a supernova model with a progenitor star of 10.6 \,M$_\odot$ and an explosion energy of 0.9 $\times$ 10$^{51}$\,erg. Approximately 15.3\% of the generated patterns were best matched by a different supernova model, featuring a higher progenitor mass of 26.0\,M$_\odot$ and a more energetic explosion of \mbox{3.0 $\times$ 10$^{51}$\,erg}. The remaining 5.8\% of the generated patterns were matched by 30 additional models. This suggests that the current abundance measurements and their associated uncertainties are not yet sufficient to fully constrain the enrichment history. Future detections of key elements, such as nitrogen, will be essential to distinguish between these models.

The predicted yields of two supernova models with roughly similar weights have ``best-fit'' results for about 88.5\% of the generated patterns for G470812. These models have slightly similar stellar masses, but different explosion energies. The remaining 11.5\% of the generated patterns were reproduced by 72 models that span a wide range of stellar masses and explosion energies. In general, these best-fitting supernova models have stellar masses spanning \mbox{$15\,M_{\odot}$} and \mbox{$30\,M_{\odot}$}, peaking at \mbox{$\approx 20\,M_{\odot}$}, which may reflect the initial mass function of the first stars.


Overall, the best-fit progenitor masses and explosion energies found for the program stars agree with the ranges presented in \citet{placco2025}, Figure 10. At [Fe/H]$\leq-3.0$ and sub-solar carbon abundances, the explosion energies would typically be $E \geq 1 \times 10^{51}$\,erg and progenitor masses \mbox{$\approx 20-30\,M_{\odot}$} \citep{Ishigaki2018}.

\subsection{Origin of the Accreted Atari Disk Star G288733}

Only G288733 exhibits disk-like kinematics, with a circular orbit (e $\sim 0.2$) that remains close to the Galactic plane (within Z$_\ensuremath{\mathrm{max}} \approx 2$\,kpc). These parameters clearly associate G288733 with the Galactic disk system. Overall, the L$_z$-E location of this star aligns with the characteristics of the so-called Atari disk \citep{Mardini2022}, largely overlapping with the metal-weak thick disk. This disk component is distinguished by its low-metallicity stellar content and a notable velocity lag when compared to the thick disk (and also the thin disk). The Atari disk is believed to have formed through the accretion of a massive dwarf galaxy into the early Galactic disk to enrich it with low-metallicity stars. The association of G288733 with the Atari disk thus indicates that it may be a relic of this ancient accretion event. 

An accretion origin of this star is further supported by its observed low strontium ([Sr/H] = $-4.08$) and barium ([Ba/H] = $-3.36$) abundances (see \citealt{SASS_stars} for more details). As shown in the ``UFD-halo separation'' plot (see the last panel of Figure~\ref{fig:abund}), G288733 occupies the same region as the ultra-faint dwarf (UFD) galaxy stars in the [Sr/Ba] versus [Ba/Fe] plane (see also panel C of Figure~6 of \citealt{Mardini24a}). This distinctive signature furthermore supports the notion that G288733 likely formed in an early dwarf galaxy that was long ago accreted by the Milky Way, such as the Atari progenitor.

\subsection{Origin of the $R$-II Star G256353}

The other astrophysically important star in our sample is G256353. It shows a significant enhancement in neutron-capture elements, with [Eu/Fe] = +1.2, and exhibits [Ba/Eu] = $-$0.64, classifying it as an $r$-II star \citep[see][]{Frebel2018}. G256353 was first identified as an RPE star by \citet{Li_2022} based on stellar parameters of \mbox{$T_{\rm eff} = 4784$\,K}, [Fe/H] $= -3.22$, and $\log g = 1.58$. These values are consistent with our derived parameters (see Section~\ref{sec:params}). To demonstrate the reliability of our abundance measurements, we compare our results with those of \citet{Li_2022} to quantify any systematic uncertainties.

\begin{figure}[!htbp]
\begin{center}
\includegraphics[clip=true,width=0.47\textwidth]{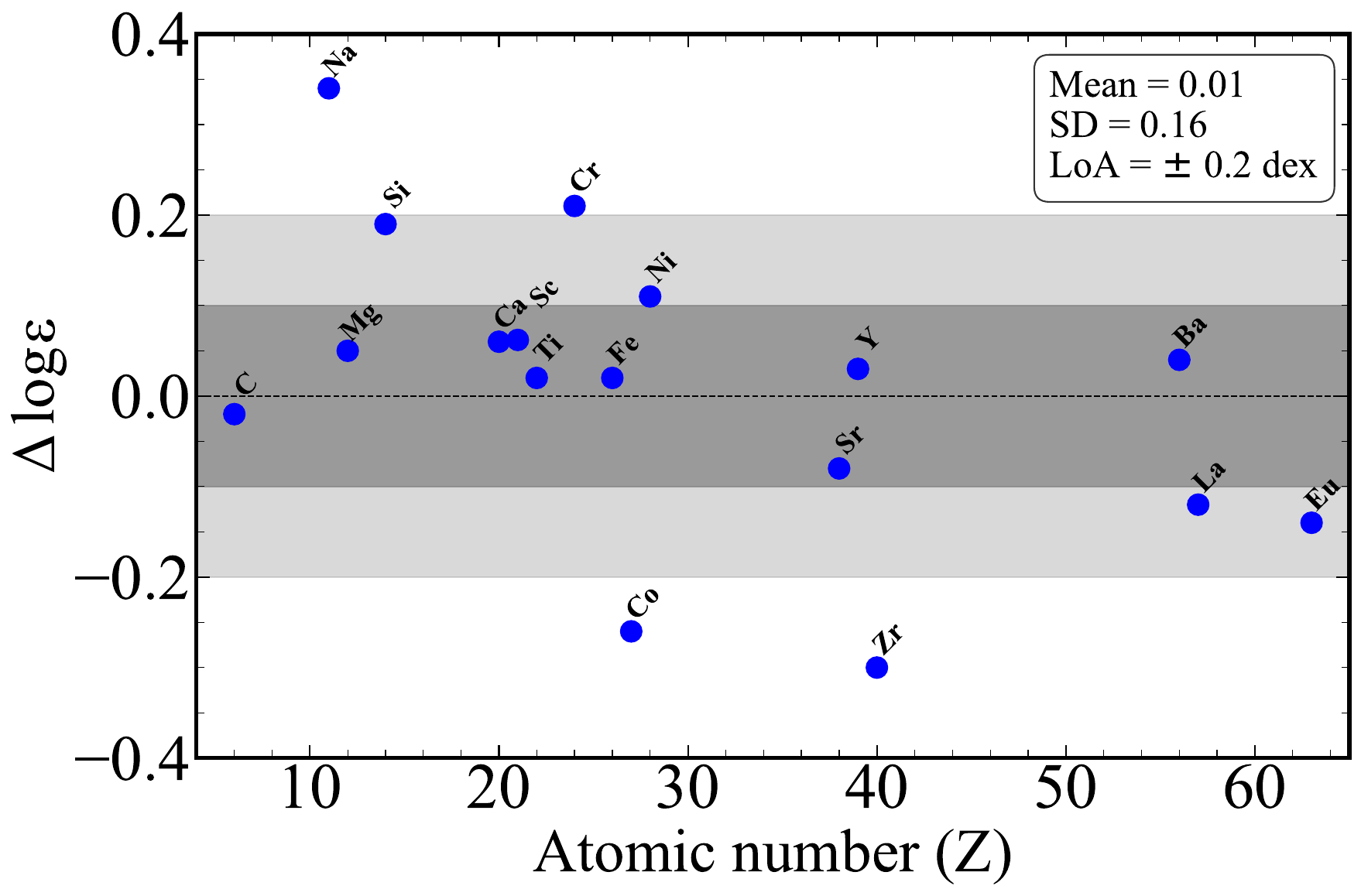} 
\caption{Comparison of elemental abundances for G256353 between this study and \citet{Li_2022}. Differences in $\log{\epsilon}(\mathrm{X})$ are plotted against atomic number, with the dashed line indicating zero and shaded bands showing $\pm0.1$ and $\pm0.2$\,dex. A Bland–Altman analysis (statistics in legend) yields a mean offset of $\sim$0.01\,dex, indicating no systematic bias. The scatter (SD $\sim$ 0.16\,dex) and limits of agreement (LoA $ \approx \pm$0.2\,dex) demonstrate overall consistency between the datasets, with minor element-to-element deviations likely due to methodological differences.
\label{fig:li_comp}}
\end{center}
\end{figure}

Figure~\ref{fig:li_comp} shows the abundance differences (as $\log{\epsilon}(\mathrm{X})$ values). To further evaluate the agreement between the two studies, we performed a Bland–Altman analysis to quantify the mean bias and the range within which most differences lie. The mean difference, standard deviation (SD), and limits of agreement (LoA) are listed in the legend of Figure~\ref{fig:li_comp}. The mean offset \mbox{of $\sim$0.01\,dex} indicates no significant systematic difference. However, a scatter of $\sim$0.16\,dex implies that individual elemental abundances may differ by a few tenths of a dex, likely reflecting methodological differences such as linelist selection, atomic data, continuum normalization, or adopting NLTE abundances. Despite the overall agreement, several elements exhibit more pronounced discrepancies: sodium shows an offset of $\Delta \log{\epsilon}(\mathrm{Na}) = +0.34$\,dex, while cobalt and zirconium differ by $-0.26$\,dex and $-0.31$\,dex, respectively\footnote{\citet{Li_2022} only provide the NLTE abundance for Na.}. Accounting for the modest differences in stellar parameters between the two studies slightly reduces these inconsistencies but does not eliminate them entirely, suggesting that residual offsets likely arise from differences in atomic data or measurement procedures.

Beyond the abundance measurements, and based on our kinematic analysis and stellar action approach (see \citealt{Mardini2022} for more details), we assign G256353 to the Galactic halo population. The star reaches a maximum vertical distance of Z$_{\ensuremath{\mathrm{max}}} \sim 6$\,kpc, placing it within the inner halo, which is generally considered to extend up to Z$_\ensuremath{\mathrm{max}} < 15$\,kpc \citep[see][and references therein]{carollo12}. The inner halo is known to host a considerable fraction of in-situ stars \citep[e.g.,][]{Bonaca17}, while the outer halo primarily contains stars with accretion histories. Based solely on the value of Z$_\ensuremath{\mathrm{max}}$   and a prograde tangential velocity of V$_{\phi}$ = 60.5\,km s$^{-1}$, an accretion origin for G256353 cannot be determined at this time, but it cannot be excluded. Further comprehensive analysis, such as identifying a potential dynamical connection between this star and any satellite $r$-process galaxies, if possible, might help to further clarify its origin.

\begin{figure*}
\begin{center}
\includegraphics[clip=true,width=\textwidth]{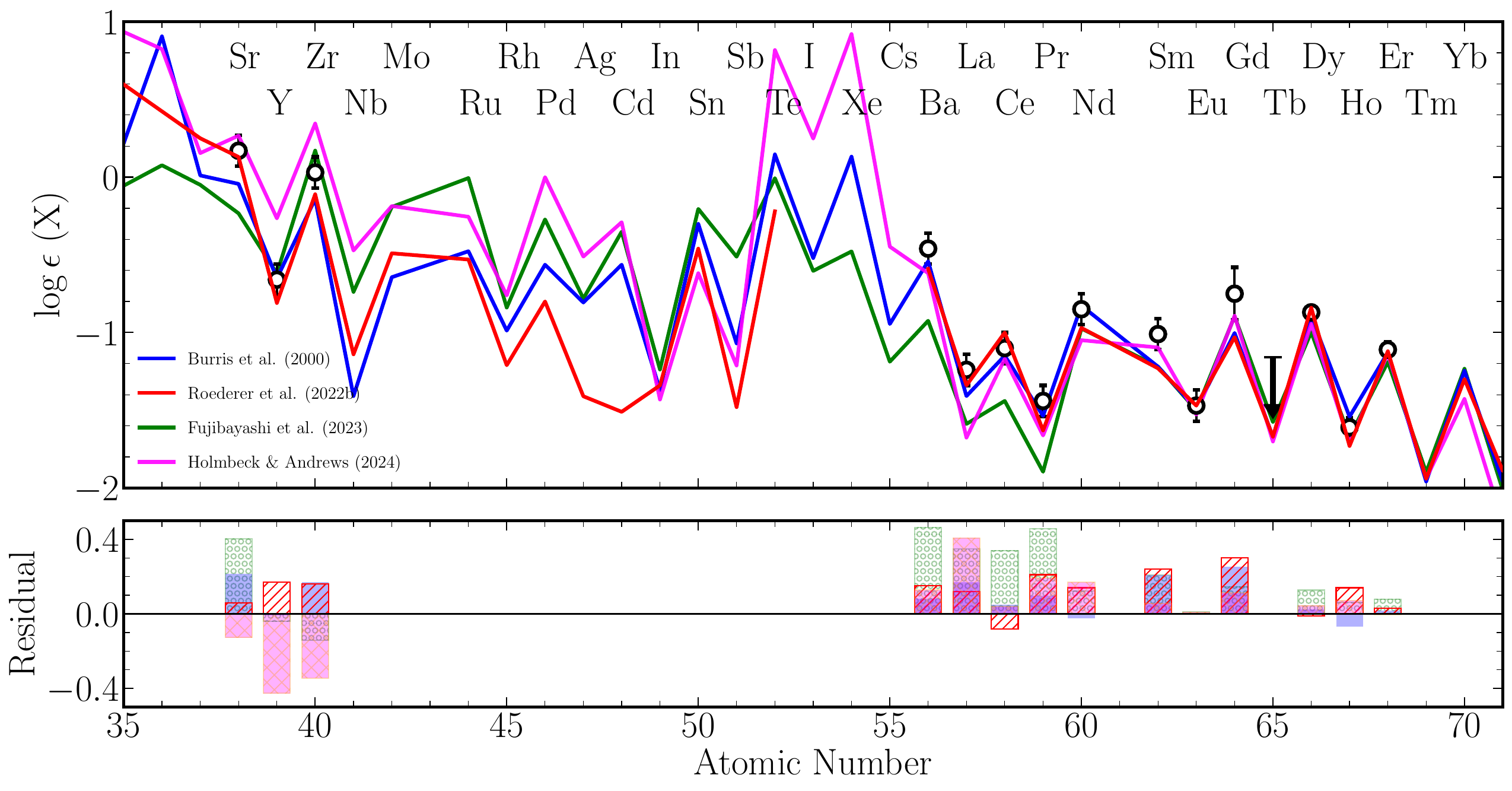} 
\caption{The neutron-capture element abundance pattern of G256353 (black open circles). We over-plot scaled $r$-process abundances of the Solar System (represented by the blue line) from \citet{burris2000}, neutron-capture element pattern of HD~222925 (shown as a red line) from \citet{Roederer2022}, the predicted yields of the SFHo-135-135 model (shown as a green line) from \citet{Fujibayashi2023}, and neutron star merger patterns which incorporate both disk winds and dynamical ejecta components from \citet{Erika_NSNS}. The patterns are scaled to match the observed Eu of G256353. Blue, dashed-red, and green bars (colors are the same as the patterns) indicate the differences between the observed abundance pattern of G256353 and the scaled patterns. The $r$-process universality is clearly seen in our star. Additionally, the scaled heavy-element pattern of HD~222925 appears slightly more consistent with that of G256353. \label{fig:r-pattern}}
\end{center}
\end{figure*}

Additional insight into the origins of G256353 in the early universe and the nucleosynthesis processes operating in its progenitor(s) can be drawn from the observed abundance pattern. Detailed simulations indicate that canonical core-collapse supernovae (CCSNe) are generally inefficient at producing third-peak $r$-process elements \citep[e.g.,][]{Qian96, Thompson2001, hudepohl2010,Voort2015}. Hence, when considering an $r$-process-enhanced star, we can assume that the astrophysical sites responsible for the lighter elements (Z $<$ 30) are likely decoupled from those that generate the heavier neutron-capture species. Accordingly, in the following discussion we consider the possibility that the observed enhancement of neutron-capture elements in G256353 originated not from a CCSNe, but from a neutron star merger.

Figure~\ref{fig:r-pattern} shows the $r$-process abundance pattern of the Solar System (blue line; \citealt{burris2000}), the benchmark RPE metal-poor star HD~222925 with the most complete stellar abundance pattern besides the Sun (red line; \citealt{Roederer2022}), and two sets of neutron star merger yield predictions (green line; \citealt{Fujibayashi2023}, and magenta line; \citealt{Erika_NSNS}), all scaled to match the observed Eu abundance of G256353. The open circles represent the average derived abundance for each element of our star, with error bars indicating the uncertainties. The lower panel of Figure~\ref{fig:r-pattern} shows the residuals between our observed abundances and the four patterns, represented by blue, red, magenta, and green bars, respectively.

Overall, the light neutron-capture element abundances (Sr, Y, and Zr) show good agreement with those measured for HD~222925 as well as the scaled Solar abundances except for Sr, which is $\sim0.2$\,dex higher in G256353. The observed abundances of the second and third $r$-process peak elements align closely with both the HD~222925 and the scaled Solar $r$-process patterns. This close agreement supports the known universality of the main $r$-process, as demonstrated across numerous RPE metal-poor stars that span a wide metallicity range \citep[e.g.,][]{Hilletal:2002,sneden2003,UII_Frebel,placco2020,Mardini_2020,Hansen2025}. Such consistency implies that a robust nucleosynthetic mechanism is responsible for producing these heavy elements that appears to be operating across diverse astrophysical environments. 

Comparing the two different theoretical neutron star merger patterns from \citet{Erika_NSNS} and \citet{Fujibayashi2023} against our observed neutron-capture element abundances reveals noticeable differences. The predicted pattern from the \citet{Erika_NSNS} model, which incorporates both disk winds and dynamical ejecta components, shows reasonable agreement with our measured abundances across both the second and third $r$-process peaks except for La, which is under-predicted by $\sim$0.4\,dex. Y and Zr, on the other hand, are significantly over-predicted. The other neutron star merger model reported by \citet{Fujibayashi2023} exhibits a similar behavior. La, Ce, and Pr, as well as Sr, are all under-produced (at a $\sim$0.3\,dex level). 

Given that the \citet{Erika_NSNS} model matches the abundance pattern of G256353 reasonably well, it suggests that it reasonably captures the nucleosynthetic conditions of heavy-element production within a neutron star merger that enriched the natal gas cloud. The mismatch between the two key elements, Sr and La, may result from uncertainties in the nuclear physics inputs (e.g., nuclear reaction rates, fission yields, entropy, gas rate of expansion, initial neutron richness of the ejecta, etc.) used in the simulations \citep[e.g.,][]{NSNS_uncertanties,NSNS_uncertanties3,NSNS_uncertanties2}. Therefore, measuring additional heavy-element abundances in this star would be helpful in further discerning the origin(s) of these elements.

In conclusion, while the observed heavy-element pattern of G256353 appears to be best described by the same pattern as HD~222925, it alone does not provide any progenitor information. However, the neutron star merger models of \citet{Erika_NSNS} and \citet{Fujibayashi2023} appear promising, despite a number of element discrepancies. More abundances of key elements, such as Nb, Pd, and Ag, would be helpful for further constraining the different model assumptions. Ultimately, the detailed abundance pattern of G256353 contributes to our growing understanding of $r$-process universality and the Galactic build-up of neutron-capture elements.

\section{Summary}\label{sec:conc}

This study presents a detailed chemodynamical analysis of an $r$-II metal-poor star G256353, characterized by [Eu/Fe] = +1.20 and \mbox{[Ba/Eu] = $-$0.64}. We selected G256353 from the Gaia BP/RP dataset (M. Mardini et al. 2026, in preparation) alongside three newly discovered EMP halo stars. High-resolution spectroscopy was obtained for all targets using the GHOST spectrograph on the Gemini-South telescope. The analysis of G256353 contributes to the ongoing efforts of the RPA to identify and characterize RPE metal-poor stars.

The chemical abundances of 30 elements, including 15 heavy elements, were derived for G256353. We compared the abundances of the light elements (Z $\leq 30$) with those predicted by Population\,III supernova models from \citet{heger_woosley10} to gain insight into the origin of these elements. This comparison suggests that a massive progenitor star (M${\star}\sim$30\,M$_{\odot}$) underwent an energetic explosion before enriching the birth cloud from which G256353 formed.

The abundance analysis of G256353 provides valuable constraints on $r$-process nucleosynthesis in the early universe. Its heavy-element pattern shows strong agreement with that of the benchmark star HD~222925. This consistency also appears when compared with the Solar $r$-process pattern for elements beyond the first peak, further supporting the universality of the main $r$-process.

Comparisons with neutron star merger models reveal notable discrepancies across the different $r$-process peaks. Although light neutron-capture elements (Sr, Y, Zr) show varying levels of agreement between different models, elements from Ba to Pr exhibit substantial over-abundances relative to the predictions from \citet{Fujibayashi2023}. These elements exhibit significantly better agreement with the multi-component model of \citet{Erika_NSNS}. In particular, the magnitude of these discrepancies is greatly reduced for elements located beyond the second $r$-process peak. These discrepancies likely originate from uncertainties in nuclear physics inputs, underscoring the need for additional abundance measurements. In particular, key elements such as Nb, Pd, and Ag would provide critical constraints on these neutron star merger models, advancing our understanding of heavy-element production and Galactic chemical evolution.

In addition to G256353, we identified and analyzed three other EMP stars from our observed sample. G297027 and G470812 appear to be typical halo EMP stars. The third star, G288733, shows a kinematic association with Atari, a metal-poor accreted structure in the Galactic disk system. The star exhibits a low abundance of strontium ([Sr/Fe] = $-$1.09) and moderate abundance of barium ([Ba/Fe] = $-$0.37), which is consistent with those reported for UFD stars. This distinctive abundance pattern of neutron-capture elements suggests an ex-situ origin for G288733, indicating that the star was most likely accreted from a smaller external system rather than forming within the Milky Way.

The data and methods applied here serve as a seed work that will set the stage for a much larger sample of $\sim200$ EMP stars observed with the MIKE spectrograph on the Magellan Telescope. The future analysis will provide homogeneous results, allowing us to delve deeper into understanding the chemical trends across different stellar populations and place crucial constraints on the formation and evolution of the Milky Way.

\begin{acknowledgments}
M.K.M. acknowledges support from NSF grant OISE 1927130 (International Research Network for Nuclear Astrophysics/IReNA).
A.F. acknowledges support from NSF-AAG grant AST-2307436. 
The work of V.M.P. is supported by NSF NOIRLab, which is managed by the Association of Universities for Research in Astronomy (AURA) under a cooperative agreement with the U.S. National Science Foundation. 
T.C.B. acknowledges partial support from grants PHY 14-30152; Physics Frontier Center/JINA Center for the Evolution of the Elements (JINA-CEE), and OISE-1927130; The International Research Network for Nuclear Astrophysics (IReNA), awarded by the US National Science Foundation, and DE-SC0023128; the Center for Nuclear Astrophysics Across Messengers (CeNAM), awarded by the U.S. Department of Energy, Office of Science, Office of Nuclear Physics. 
T.T.H. acknowledges support from the Swedish Research Council (VR 2021-05556).
I.U.R.\ acknowledges support from the U.S.\ National Science Foundation (grant AST~2205847).

Based on observations obtained at the International Gemini Observatory (Program ID: GS-2023B-FT-301), a program of NSF NOIRLab, which is managed by the Association of Universities for Research in Astronomy (AURA) under a cooperative agreement with the U.S. National Science Foundation on behalf of the Gemini Observatory partnership: the U.S. National Science Foundation (United States), National Research Council (Canada), Agencia Nacional de Investigaci\'on y Desarrollo (Chile), Ministerio de Ciencia, Tecnolog\'ia e Innovaci\'on (Argentina), Minist\'erio da Ci\^encia, Tecnologia, Inova\c{c}\~oes e Comunica\c{c}\~oes (Brazil), and Korea Astronomy and Space Science Institute (Republic of Korea).
E.M.H.\ acknowledges this work performed under the auspices of the U.S.\ Department of Energy by Lawrence Livermore National Laboratory under Contract DE-AC52-07NA27344 with release number LLNL-JRNL-2014233.
A.F.\ acknowledges support from NSF-AAG grant AST-2307436.
I.U.R.\ acknowledges
support from NSF grant AST~2205847 and the NASA Astrophysics Data Analysis Program, grant 80NSSC21K0627.
T.T.H\ acknowledges support from the Swedish Research Council (VR 2021-05556).
\end{acknowledgments}

This work has made use of data from the European Space Agency (ESA) mission
{\it Gaia} (\url{https://www.cosmos.esa.int/gaia}), processed by the {\it Gaia}
Data Processing and Analysis Consortium (DPAC,
\url{https://www.cosmos.esa.int/web/gaia/dpac/consortium}). Funding for the DPAC
has been provided by national institutions, in particular the institutions
participating in the {\it Gaia} Multilateral Agreement.
This job has made use of the Python package GaiaXPy, developed and maintained by members of the Gaia Data Processing and Analysis Consortium (DPAC), and in particular, Coordination Unit 5 (CU5), and the Data Processing Centre located at the Institute of Astronomy, Cambridge, UK (DPCI).
This research has made use of the SIMBAD database,
operated at CDS, Strasbourg, France.

\software{
{\texttt{The-ORIENT}}\,\citep[\url{https://github.com/Mohammad-Mardini/The-ORIENT};][]{Mardini_2020,Mardini2022b}, 
{\texttt{linemake}}\,\citep[\url{https://github.com/vmplacco/linemake/};][]{Placco2021_linemake,Placco2021ascl}
{\texttt{IRAF}}\,\citep[\url{https://iraf.noirlab.edu/};][]{tody1986,tody1993,fitzpatrick2025}, 
{\texttt{DRAGONS}}\,\citep[\url{https://dragons.readthedocs.io/en/release-3.2.x/};][]{labrie2023,dragons}, 
{\texttt{SMHr}}\,\citep[\url{https://github.com/andycasey/smhr};][]{casey14},
{\texttt{GaiaXPy 1.1.3}}\,\citep[\url{https://gaia-dpci.github.io/GaiaXPy-website/};][]{casey14},
{\texttt{numpy} \citep[\url{https://numpy.org;}][]{VanDerWalt_2011_numpy, Harris_2020_numpy}, 
{\texttt{matplotlib} \citep[\url{https://matplotlib.org};][]{Hunter_2007_matplotlib}, 
{\texttt{scipy} \citep[\url{https://scipy.org}; \href{https://mail.python.org/pipermail/python-list/2001-August/106419.html}{Jones et al. 2001};][]{Virtanen_2020_scipy}, 
{\texttt{astropy} \citep[\url{http://www.astropy.org};][]{AstropyCollaboration_2013, AstropyCollaboration_2018, AstropyCollaboration_2022}
}}}}}

\facilities{
Gemini:South (GHOST)
}

\appendix

\section{Compilation references for Table~2}
(1) NIST \cite{kramida2020}; (2) \citet{pehlivan2017}; (3) NIST \cite{kramida2020} for log($gf$) value and VALD \citet{piskunov1995,pakhomov2019} for HFS; (4) \citet{trabert1999} for log($gf$) value and \citet{roederer2021} for HFS;
(5) \citet{denhartog2023}; (6) \citet{denhartog2021b}; (7) \citet{lawler1989} for $\log g$ and \citet{kurucz2011} for HFS; (8) \citet{lawler2019};
(9) \citet{lawler2013}; (10) \citet{wood2013}; (11) \citet{pickering2001}, with corrections given in \citet{pickering2002}; (12) \citet{wood2014a} for log($gf$) value and HFS; (13) \citet{sobeck2007}, (14) \citet{lawler2017}, (15) \citet{gurell2010}; (16) \citet{nilsson2006}; (17) \citet{denhartog2011} for both log($gf$) value and hfs; (18) \citet{obrian1991}; (19) \citet{belmonte2017}; (20) \citet{denhartog2014}; (21) \citet{ruffoni2014}; (22) \citet{denhartog2019}; (23) \citet{melendez2009}; (24) \citet{lawler2015} for log($gf$) values and HFS; (25) \citet{lawler2018} for log($gf$) value and \citet{roederer2022a}for HFS; (26) \citet{wood2014b}; (27) \citet{fedchak1999}; (28) \citet{roedererlawler2012}; (29) \citet{biemont2011}; (30) \citet{ljung2006}; (31) \citet{malcheva2006}; (32) \citet{nilsson2010}, including HFS; (33) \citet{sikstrom2001}; (34) \citet{xu2004}; (35) \citet{kramida2018}, using HFS/IS from \citet{mcwilliam1998}; (36) \citet{lawler2001a}, using HFS from \citet{ivans2006} when available; (37) \citet{lawler2009}; (38) \citet{li2007}, using HFS from \citet{sneden2009}; (39) \citet{denhartog2003}, using HFS/IS from \citet{roederer2008} when available; (40) \citet{lawler2006}, using HFS/IS from \citet{roederer2008} when available; (41) \citet{lawler2001c}, using HFS/IS from \citet{ivans2006}; (42) \citet{denhartog2006}, (43) \citet{lawler2001b}, using HFS from \citet{lawler2001d}; (44) \citet{wickliffe2000}; (45) \citet{lawler2004}, using HFS from \citet{lawler2009}; (46) \citet{lawler2008}; (47) \citet{wickliffe1997}; (48) \citet{sneden2009} for log($gf$) value and HFS/IS; (49) \citet{roederer2010} for log($gf$) value and \citet{denhartog2020} for HFS; (50) \citet{denhartog2021a}; (51) \citet{lawler2007}; (52) \citet{denhartog2005} for log($gf$) value and HFS/IS; (54) \citet{denhartog2005} for log($gf$) value only; (55) \citet{hannaford1981} for log($gf$) value and \citet{demidov2021}for HFS; (56) \citet{nilsson2002}

\section{Compilation references for Figure~2}
Metal-poor star data used for generating Figure~\ref{fig:abund} are taken from:
\citet{Aguado2017b,Aguado2017, Aguado2018,Aguado2021}; \citet{Almusleh2021}; \citet{Aoki2007, Aoki2013}; \citet{barklem05}; \citet{Behara2010}; \citet{Bonifacio2012}; \citet{Caffau2011a,Caffau2011,Caffau2013};  \citet{Carretta2002}; \citet{Casey2015}; \citet{cayrel2004}; \citet{Christlieb2002};  \citet{Cohen2004, Cohen2013}; \citet{Depagne2000}; \citet{ezzeddine19}; \citet{For2010}; \citet{francois07}; \citet{Frebel2010, frebel08,frebel15b, Frebel2019, HE1327_Nature}; \citet{Hansen2015}; \citet{Hansen25}; \citet{Hollek2011}; \citet{Honda2011}; \citet{Jacobson2015}; \citet{Keller2014}; \citet{Lai2004}; \citet{Lai2008}; \citet{Limberg25}; \citet{Mardini_2019a,Mardini_2019b,Mardini_2020,Mardini2022b}; \citet{Masseron2006}; \citet{Norris2001, Norris2007}; \citet{Ou25}; \citet{placco2014a,placco2020};  \citet{Plez2005}; \citet{Rich2009}; \citet{roederer14c}; \citet{Ryan1991}; \citet{Ryan1996}; \citet{Ryan1999}; \citet{Sivarani2006}; \citet{Spite1999}; \citet{Spite2000}; \citet{Spite2014}; \citet{Yong2013}.

\bibliography{references, Terese}{}
\bibliographystyle{style/aasjournal}

\end{document}